\documentclass[aps,twocolumn, superscriptaddress,table]{revtex4-2}
\usepackage{multirow} 
\usepackage[english]{babel}
\usepackage[T1]{fontenc} 
\usepackage{lmodern}
\usepackage{graphicx}
\usepackage[colorinlistoftodos, color=green!40, prependcaption]{todonotes}

\usepackage{tabularx}

\usepackage[normalem]{ulem}
\useunder{\uline}{\ul}{}

\makeatletter
\def\maketitle{
\@author@finish
\title@column\titleblock@produce
\suppressfloats[t]}
\makeatother

\usepackage{multibib}

\def \IFD{Institute of Experimental Physics, Faculty of Physics, University of Warsaw, Pasteura St. 5, 02-093 Warsaw, Poland}
\def \IFT{Institute of Theoretical Physics, Faculty of Physics, University of Warsaw, Pasteura St. 5, 02-093 Warsaw, Poland}
\def \Konst{Department of Physics, University of Konstanz, Universit{\"a}tsstra{\ss}e 10, D-78457 Konstanz, Germany}
\def \PW{Department of Semiconductor Materials Engineering Faculty of Fundamental Problems of Technology, Wrocław University of Science and Technology Wybrzeże Wyspiańskiego 27, 50-370 Wrocław, Poland}
\def \HaifaChem{Schulich Faculty of Chemistry, Technion – Israel Institute of Technology, Haifa 3200003, Israel}
\def \HaifaMat{Department of Materials Science and Engineering, Technion – Israel Institute of Technology, Haifa 3200003, Israel}
\def \AGH{Faculty of Physics and Applied Computer Science, AGH University of Kraków, Mickiewicza 30, 30-059 Kraków, Poland}
\def \CeNT{Chemical and Biological Systems Simulation Lab, Centre of New Technologies, University of Warsaw, 02-097 Warsaw, Poland}

\graphicspath{{./Images/}}

\begin{document}

\title{Direct Optical Probing of the Magnetic Properties\\ of the Layered Antiferromagnet CrPS$_4$}

\author{Tomasz~Fąs}\affiliation{\IFD}
\author{Mateusz~Wlazło}\affiliation{\CeNT}
\author{Magdalena~Birowska}\affiliation{\IFT}
\author{Miłosz~Rybak}\affiliation{\PW}
\author{Małgorzata Zinkiewicz}\affiliation{\IFD}
\author{Leon Oleschko}\affiliation{\Konst}
\author{Mateusz Goryca}\affiliation{\IFD}
\author{Łukasz Gondek}\affiliation{\AGH}
\author{Bruno~Camargo}\affiliation{\IFD}
\author{Jacek~Szczytko}\affiliation{\IFD}
\author{Adam~K.~Budniak}\affiliation{\HaifaChem}
\author{Yaron~Amouyal}\affiliation{\HaifaMat}
\author{Efrat~Lifshitz}\affiliation{\HaifaChem}
\author{Jan~Suffczyński}\email[*]{jsuffczynski@uw.edu.pl}\affiliation{\IFD}

%\email[Correspondence email address: ]{j.suffczynski@uw.edu.pl}% Your name

\date{\today} % Leave empty to omit a date

\begin{abstract}
Unusual magnetic properties of Van der Waals type antiferromagnetic semiconductors make them highly attractive for spintronics and optoelectronics. A link between the magnetic and optical properties of those materials, required for practical applications, has not been, however, established so far. Here, we report on a combined experimental and theoretical study of magnetic, optical, and structural properties of bulk CrPS$_{4}$ samples. We find that the magnetic-field-dependent circular polarization degree of the photoluminescence is a direct measure of the net magnetization of CrPS$_{4}$. Complementary, Raman scattering measured as a function of magnetic field and temperature enables the determination of the magnetic susceptibility curve of the material. Our experimental results are backed by Our experimental results are supported by density functional theory calculations that take as input the lattice parameters determined from temperature-dependent X-ray diffraction measurements. This allows us to explain the impact of spin ordering on the spectral position of Raman transitions in CrPS$_4$, as well as anomalous temperature shifts of selected of them. The presented method for all-optical determination of the magnetic properties is highly promising for studies of spin ordering and magnetic phase transitions in single- or a few-layer samples of magnetic layered materials, for which a poor signal-to-noise ratio precludes any reliable neutron scattering or magnetometry measurements. 

\end{abstract}


\maketitle

%\section{Introduction}

2D materials, such as binary transition metal dichalcogenides (TMDs), have recently been at the forefront of materials research.\cite{Choi:MatToday2017,Pacuski_NL_2020} TMDs offer spin and valley degrees of freedom, but in general, lack magnetism, which limits their potential for applications in spintronic or optoelectronic devices. % Substitutional doping of TMDs with magnetic ions is limited to the level as low as a few molar percent.\cite{Tedstone:ChemofMat2016} 
An alternative is to consider 2D semiconducting materials whose lattice natively incorporates transition metal ions with a non-vanishing spin moment such as Mn, Ni, Cr, or Fe.\cite{Burch:Nature2018, Gong:Science2019, Rahman:ACSNano2021, Autieri2022, Hu:NatSciOpen2023} The prominent examples include transition metal phosphorus chalcogenides such, as NiPS$_3$, MnPS$_{3}$, FePS$_{3}$, or CrPS$_{4}$.\cite{Wang:AdvFunMat2018, Qian:JPhysChemC2020} This class of materials exhibits an antiferromagnetic ordering with N\'{e}el temperature (T$_N$) of up to 150~K. A variety of their potential applications in electronics and optospintronics include spin tunnel field-effect transistors or electrical switches with carrier mobility controlled by strain.\cite{Chittari:PRB2016, Jiang:NatureElectr2019, Rahman:ACSNano2021, Hu:NatSciOpen2023} The intrinsic robustness of 2D antiferromagnetic materials  against perturbations induced by an external magnetic field, and the possibility of controlled directing of the N\'{e}el vector along one of the (typically multiple) magnetic easy axes, are advantageous for their implementation in non-volatile memory devices.

In the present work, we investigate CrPS$_{4}$ (chromium phosphorus tetrasulphide or chromium thiophosphate), which has been the focus of increasing attention. It is a rare case of a layered A-AFM-type antiferromagnet, where magnetic ions (Cr$^{3+}$) within van der Waals planes are organised in a rectangular lattice. CrPS$_{4}$ is indicated as a promising material for spintronic, electrocatalytic, energy-storage, optoelectronic and quantum information applications\cite{Mayorga:ASCAppMat2017,Joe:JOP2017, Gu:ASCNano2020, Riesner:JChemPhys2022, Ding_2020} or even for the generation of a tunable X-ray emission.\cite{Shentcis:NatPhotonics2020}

Despite the increasing number of studies on CrPS$_4$,\cite{Zhuang:PRB2016,Lee:ACSNano2017,Wu:RSCAdv2019,Kim:NanoLett2019, Budniak:Small2020, Shentcis:NatPhotonics2020, Ding_2020, Peng:AdvMat2020,Susilo:npjQMater2020} a link between its magnetic phase and optical response has not been established so far. Here, we address this issue by choosing photoluminescence (PL) and Raman scattering as our optical spectroscopy tools. Both methods have been proven efficient in studies of spin ordering in conventional bulk\cite{Kluczyk:arxive2023, Grzybowski:Nanoscale2024} and layered 2D antiferromagnetic materials.\cite{Kim:2DMaterials2019, McCreary:NatureComm2020, Wang:NatureMat2021, Zhang:NanoLett2021, Vaclavkova:PRB2021} The usefulness of the PL stems from the fact that in many cases the PL signal originates from internal transitions of the magnetic ions, thus carries the information of the magnetic state of those ions. As for the Raman scattering, an exchange-type term contributes to interaction energy between the magnetic cations, which affects the vibrational characteristics of the lattice. As a result, at least some of the phonon modes are affected by the type of spin ordering in the crystal lattice and magnetic phase transitions. Thanks to strong phonon interactions and large exciton binding energy in 2D materials, even a few-layer samples provide a sizeable Raman scattering and PL signal. Spectroscopy of Raman- or PL-type is particularly useful in the case of \textit{antiferromagnets} that inherently provide only weak Kerr-rotation or magnetic circular dichroism signal.\cite{Son:ACSNano2021}

Hitherto, PL and Raman spectroscopy studies of CrPS$_{4}$ were conducted in the absence of magnetic field. They concentrated on layer thickness and polarization dependence of the Raman signal,\cite{Lee:ACSNano2017, Kim:NanoLett2019, Wu:RSCAdv2019, Kim:JPCh2021} the issue of the material stability,\cite{Kim:NanoLett2019, Son:ACSNano2021} and on the origin of the PL signal and its pressure dependence.\cite{Susilo:npjQMater2020, Riesner:JChemPhys2022, Kim:ACSNano2022, Peng:AdvFunctMat2022} As a result, little is known about magneto-optical properties of this material.

%To address this issue, here we combine
In the present work, to uncover a link between the magnetic and optical properties of the CrPS$_{4}$, we combine optical magneto-spectroscopy of bulk and thin layer samples of CrPS$_{4}$ with SQUID magnetometry. We support the interpretation of the experimental data by density functional theory (DFT) modelling of the CrPS$_{4}$ phonon structure taking as an input lattice parameters provided by temperature-dependent X-ray diffraction (XRD).

With the magnetic field and temperature-dependent experiments, we establish that the circular polarization degree of the PL provides a direct probe of the magnetization of CrPS$_4$, while Raman scattering enables the determination of its magnetic susceptibility curve. The calculations taking into account the spin ordering of Cr$^{3+}$ cations clarify an energy order of different magnetic phases of CrPS$_{4}$, explain their impact on a spectral position of Raman transitions, and describe the observed anomalous spectral shifts with temperature. As such, our work provides a comprehensive picture of structural, magnetic, optical and electronic properties of CrPS$_{4}$. We envision that the presented magneto-optical method for determining magnetic susceptibility curve and magnetic-phase transitions will be highly useful in the case of very thin samples of magnetic materials down to a single layer level, for which insufficient signal-to-noise ratio precludes the employment of standard methods to determine magnetic ordering such as SQUID magnetometry or neutron scattering.

\section{Results and discussion}
\subsection{Structural characterization of the CrPS$_{4}$ samples}

CrPS$_{4}$ single crystals are synthesized from stoichiometric amounts of elements via physical vapor transport without any transport agent (see Methods).\cite{Budniak:Small2020}

\begin{figure}[h!]
\centering
\includegraphics[width=\columnwidth]{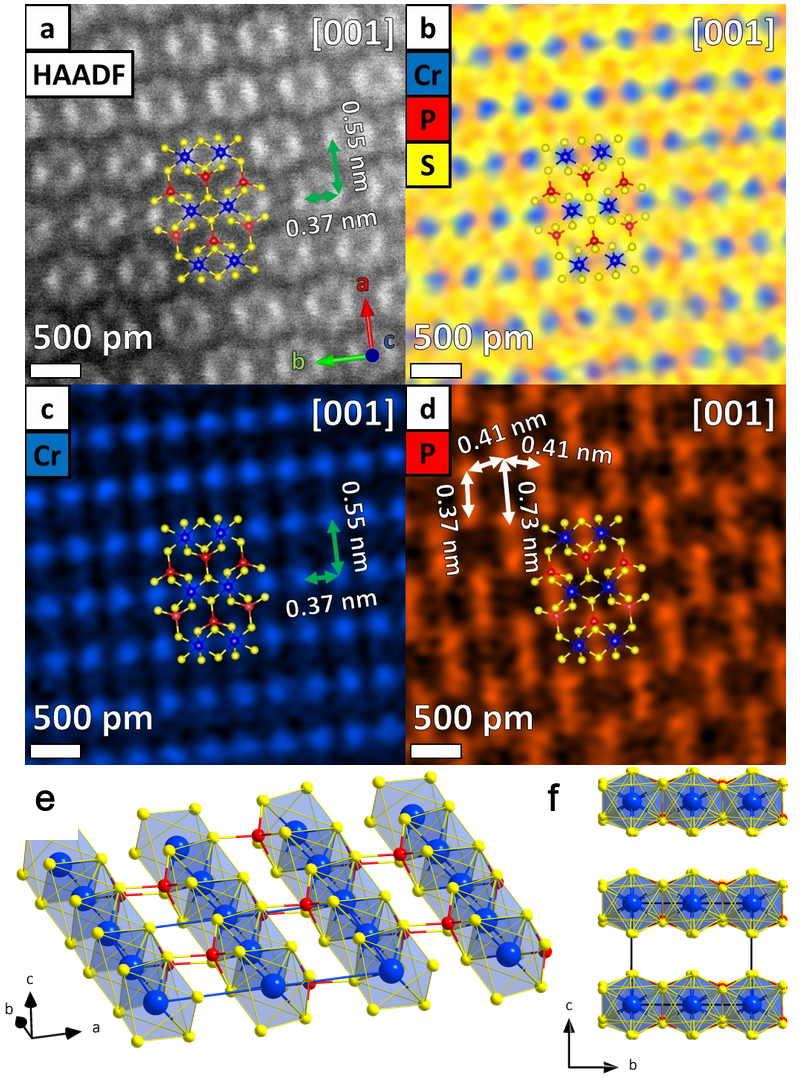}% 
\caption{High-resolution scanning transmission electron microscopy with atomic resolution of a mechanically exfoliated CrPS$_{4}$ layer. (a) HAADF-STEM micrograph. EDS elemental maps presenting the atomic positions of (b) Cr, P, and S, (c) Cr only, (d) P, and S. The micrographs are overlapped with simulated structures showing the position of the atom columns of chromium (blue spheres), phosphorus (red spheres), and sulfur (yellow spheres). Scale bars are indicated. A schematic view of a CrPS$_{4}$ layered structure: (e) a single layer in $a-b$ crystal plane with apparent chain-like atomic structures, (f) a side view of $a-b$ planes bound by van der Waals forces along the $c$ axis.}
 \label{fig:HRSTEM}
\end{figure}

\begin{figure*}[ht!]
%\centering
 \includegraphics[width=\linewidth]{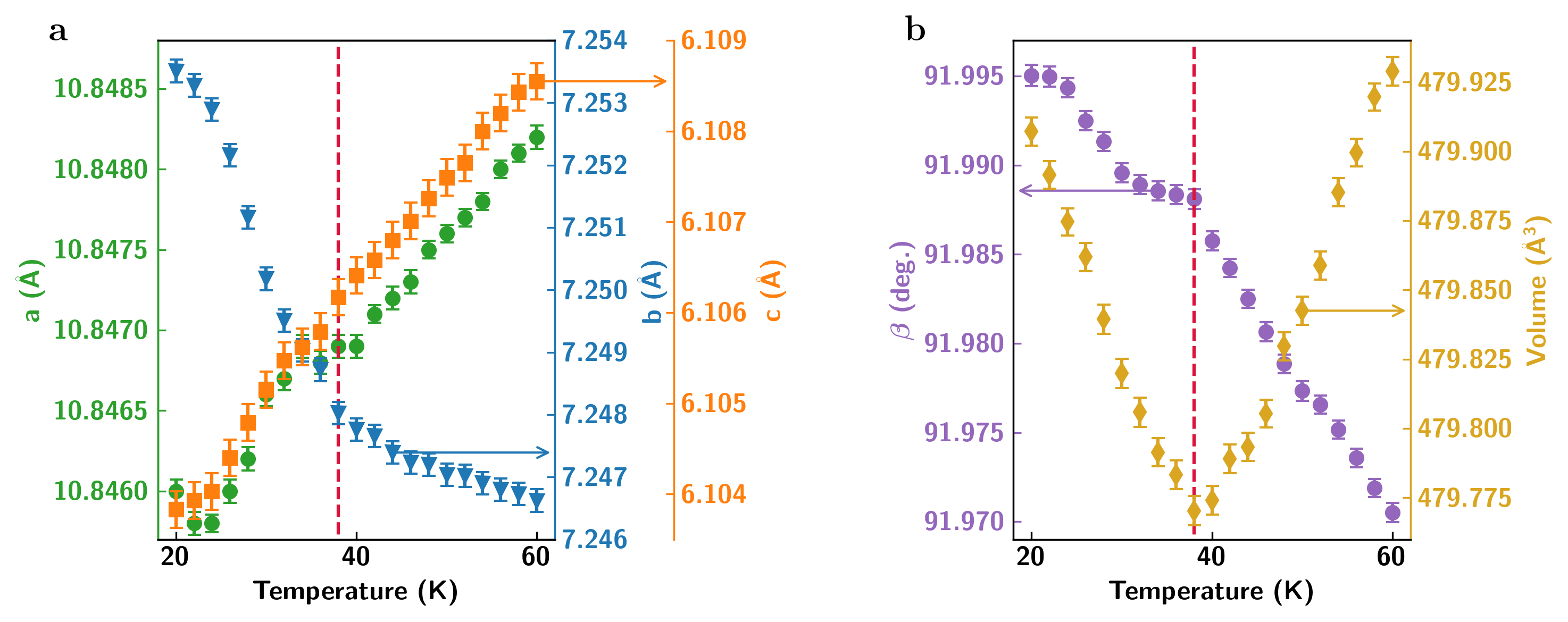}
\caption{(a) Lattice parameters ($a, b$, and $c$), (b) monoclinic angle and $\beta$ and unit cell volume as a function of temperature determined by non-ambient X-ray diffraction studies conducted on powdered samples of CrPS$_{4}$. An anomalous-type dependence of the $b$ lattice parameter is observed. A minimum in the dependence of the cell volume clearly marks position of the N\'{e}el temperature at 38~K.}
 \label{fig:XRD}
\end{figure*}

To perform a structural characterization of the CrPS$_{4}$ samples, we employ a range of methods, including an aberration-corrected scanning transmission electron microscopy (STEM) with high-angle annular dark field (HAADF) investigation and energy-dispersive X-ray spectroscopy (EDS), temperature dependent X-ray diffraction, as well as scanning electron microscopy (SEM). Fig.~\ref{fig:HRSTEM}a presents a micrograph registered in [001] zone axis of a mechanically exfoliated CrPS$_{4}$ crystal using a HAADF detector.

Fig.~\ref{fig:HRSTEM}a shows a HAADF micrograph registered in the [001] zone axis of a mechanically exfoliated CrPS$_{4}$ crystal. A molecular model\cite{Louisy:SSC1978} is superimposed on a micrograph, with the chromium atoms (shown by blue spheres) form a rectangular lattice in the $a-b$ plane. Distances between the Cr atoms along the $a$ and $b$ axes are determined to be 0.55~$\pm$~0.02~nm and 0.37~$\pm$~0.02~nm, respectively. Phosphorus atoms ($Z = 15$) cannot be distinguished from sulfur atoms ($Z = 16$) by the HAADF detector because of the too small atomic number difference. In order to resolve signals related to these two atoms, we switch to atomically-resolved EDS mapping. The result of the mapping is shown in Fig.~\ref{fig:HRSTEM}b, where sulfur atoms are represented by yellow and phosphorus atoms with red spheres. Fig.~\ref{fig:HRSTEM}c shows the result of EDS mapping of the chromium atoms only. The distances between rectangularly arranged columns, 0.55~$\pm$~0.02~nm and 0.37~$\pm$~0.02~nm, are in full agreement with the result of the HAADF measurement. Additionally, Fig.~\ref{fig:HRSTEM}d presents the EDS map of phosphorus atoms. The distances between respective pairs of them are 0.37~$\pm$~0.02~nm, 0.41~$\pm$~0.02~nm, and 0.73~$\pm$~0.02~nm, and agree with previous works.\cite{Budniak:Small2020}

Fig.~\ref{fig:HRSTEM}e shows a schematic view of a single layer of CrPS$_4$ in $a-b$ plane. Sulfur atoms in CrPS$_{4}$ are located in puckered layers, while chromium atoms are coordinated in the middle of distorted octahedrons formed by six sulfur atoms. Phosphorus atoms are in the middle of tetrahedrons of four sulfur atoms. Since these phosphorus-sulfur tetrahedrons are placed between columns of chromium-sulfur octahedrons aligned along the $b$ axis, chromium-chromium distances are longer along the $a$ axis than the $b$ axis, which is the source of a structural in-plane anisotropy of the CrPS$_4$ crystal. The van der Waals gap lies across $a-b$ planes and layers forming CrPS$_{4}$ are stacked along the $c$ direction, as shown in Fig.~\ref{fig:HRSTEM}f.

XRD measurements on powdered samples are conducted as a function of temperature in the range between 20~K and 300~K. XRD patterns reveal that in the investigated temperature range all reflections can be indexed by the monoclinic \textit{C2} (No. 5) space group. The measurements reveal complex dependencies, as depicted in Fig.~\ref{fig:XRD}a. Firstly, the lattice parameter $b$ shows an anomalous decrease with the rise of the temperature. The decrease is sudden below 38~K and slows down above this temperature, reaching a minimum at 140~K (see Suppl.~Fig.~\ref{fig:XRD_suplement} in Supplementary Information). Lattice parameters $a$ and $c$ rise according to a typical thermal expansion of the lattice, with a slight increase in the expansion rate around the T$_N$. The monoclinic angle $\beta$ between $b$ and $c$ crystallographic directions exhibits a monotonic decrease with temperature in the entire investigated range. In the vicinity of 38~K the change of the $\beta$ first slows down and then undergoes a sudden drop (see Fig.~\ref{fig:XRD}b). The dependence of the unit cell volume on temperature shows a distinct minimum at 38~K, with a roughly linear temperature dependence in the vicinity of this anomaly, as apparent from Fig.~\ref{fig:XRD}b and Suppl.~Fig.~\ref{fig:XRD_suplement}.

\begin{figure*}[ht!]
\centering
\includegraphics[width=\linewidth]{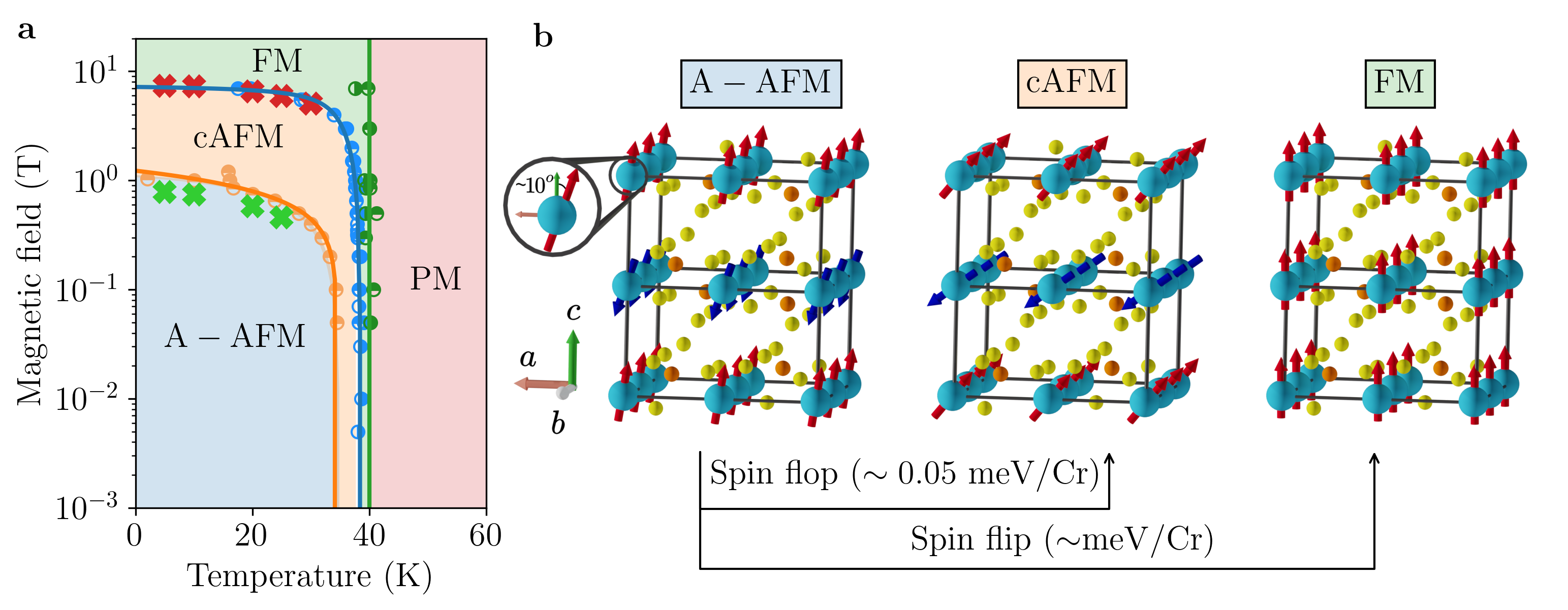}
\caption{(a) Magnetic phase diagram of CrPS$_4$ constructed from SQUID and photoluminescence data. Horizontally and vertically filled symbols represent points obtained from magnetometry measurements with the magnetic field parallel and perpendicular to the $c$-axis of the sample, respectively (see text). Crosses represent critical fields of spin-flop and spin-flip transitions determined for consecutive temperatures from the dependence of the degree of circular polarization of photoluminescence on the magnetic field parallel to the $c$-axis. Solid lines show a fit to the points from magnetometry and indicate a boundary between magnetic phases. (b) A scheme of Cr$^{3+}$ ions spin arrangements in the three magnetic phases of the lowest energy: A-AFM (a~magnetic ground state), canted AFM (cAFM) and ferromagnetic (FM) phase. A calculated difference of energy per Cr$^{3+}$ ion with respect to A-AFM phase for cAFM and FM phase is indicated. Cr, P, and S atoms are represented by blue, orange and yellow spheres, respectively. 
}
\label{fig:magn_phase_diagram}
\end{figure*}

Theoretical considerations allow us to explain a strongly anisotropic thermal expansion of the CrPS$_4$, manifested by, e.g., an anomalous-type variation of the $b$ lattice parameter with temperature. We ascribe the origin of the anisotropy to an interplay of magnetostriction, lattice dynamics and/or crystal electric field effects. The minimum of the cell volume observed at T$_N$ results from a competition in minimizing the overall energy of the crystal between Cr-related magnetic interactions (that decrease and vanish upon approaching the T$_N$) and the elastic energy of the lattice (that increases with the temperature).\cite{Gillard:NPJ2D:2024} The crystallographic direction along the $b$ axis is distinguished by the ferromagnetic interaction as being the strongest between the nearest neighbouring Cr$^{3+}$ cations along that direction. In consequence, chains of Cr$^{3+}$ ions along $b$-axis exhibit a higher stiffness than the rows of Cr$^{3+}$ ions in other crystal directions. Calculations indicate that the switching from the A-AFM phase to any different AFM type ordering shortens the lattice parameter $b$ and elongates the lattice parameters $a$ and $c$. Since the probability of any other magnetic arrangement than the ground state A-AFM increases with the rise of the temperature, the decrease of the $b$ lattice parameter is observed. Suppl.~Fig.~\ref{supplfig:ablattices_theor} shows the calculated temperature variation of the $a$ and $b$ crystal parameters, remaining in qualitative agreement with the experiment.

Results of a further structural characterization of the samples are provided in Sec.~\ref{SI:struct characterization} of Supplementary Information.

%, similarly as it was observed previously in the case of layered antiferromagnet FePS$_3$.\ref{Murayama:JAP2016}

\subsection{Magnetometry characterization of the CrPS$_{4}$ samples}

The magnetic susceptibility of CrPS$_{4}$ determined as a function of temperature for magnetic field parallel and perpendicular to the sample $c$-axis is presented respectively in Suppl.~Fig.~\ref{fig:sus}a~and~\ref{fig:sus}b. Both geometries exhibit a clear maximum at the N\'{e}el temperature of 38~K, in agreement with previous reports.\cite{Louisy:SSC1978, Peng:AdvMat2020} The position of the maximum keeps the value of 38~K (N\'{e}el temperature) for low magnetic fields ($\leq 1$~T). It is then progressively suppressed by increasing magnetic fields, reaching 19~K at 7~T. We observe an additional, smaller peak below the main maximum of the dependence in the parallel configuration, which we attribute to a phase transition (spin flop) between the A-AFM and canted AFM phase.\cite{Peng:AdvMat2020} For temperatures higher than 60~K, both parallel and perpendicular configurations exhibit a typical Curie-Weiss behaviour with the paramagnetic Curie temperature of $\theta = 40.0 \pm 0.3$~K, in agreement with previous reports.\cite{Peng:AdvMat2020, Ding_2020, Son:ACSNano2021} The positive value of $\theta$ indicates a ferromagnetic coupling between Cr$^{3+}$ spins in the $a - b$ plane.

A phase diagram of CrPS$_{4}$ constructed based on the magnetic susceptibility data collected in both, parallel and perpendicular configurations is presented respectively in Fig.~\ref{fig:magn_phase_diagram}a and ~\ref{fig:magn_phase_diagram}b. Blue cricles represent the position of the main peak of susceptibility in Suppl. Fig.~\ref{fig:sus}, while the green ones come from approximating the high temperature (100~K - 280~K) dependence by Curie-Weiss law. Orange circles represent the position of the spin flop-related peak in the susceptibility curve, defining a boundary of the transition between the AFM and the cAFM phase upon the increase of temperature or magnetic field. The boundary is described by the phenomenological relation: $H_{SF}(T) = H_{0} \left[ 1 - \frac{T}{T_{SF}} \right]^{1/2}$. The fitting to the experimental data returns the values of spin-flop temperature $T_{SF} = 34.5 \pm 0.2$~K and $H_0 = 1.23 \pm 0.05$~T. Upon further increase of the temperature, cAFM to FM and, ferro- to paramagnetic (PM) phase transitions occur. 
%s at the N\'{e}el temperature of $T_N = 38.0 \pm 0.3$~K

In order to support our interpretation of the magnetometry data, we perform DFT+U calculations (U~=~3~eV, see Methods) for various magnetic arrangements of the Cr$^{3+}$ spins. Three of the magnetic phases of CrPS$_4$ considered in the calculations are shown in Fig.~\ref{fig:magn_phase_diagram}b, while a full set of them is presented in Suppl.~Fig.~\ref{Supplfig:spinarr}. Our calculations show that the magnetic ground state of bulk CrPS$_{4}$ is A-AFM (with intralayer FM-type and interlayer AFM-type coupling of Cr$^{3+}$ spins; see Fig.~\ref{fig:magn_phase_diagram}b), which is in line with previous reports.\cite{Zhuang:PRB2016, Ding_2020, Peng:AdvMat2020} In view of our calculations, Cr$^{3+}$ spins are tilted about 10$^o$ from the $c$-axis (see Suppl.~Figure~\ref{Supplfig:Supp_exch}), with a non-zero component along the $a$ direction, in agreement with previous neutron diffraction measurements.\cite{Calder:PhysRevB2020, Peng:AdvMat2020} Spin-flop transition that rotates the Cr$^{3+}$ spins from the out-of-plane to the in-plane direction (cAFM phase) requires an energy of around 0.05~meV per magnetic ion (see Suppl.~Fig.~\ref{Supplfig:Supp_exch}). The spin-flip of the Cr$^{3+}$ spins resulting in FM phase of the crystal enforces an energy of around 1~meV. These generally small energy differences between the metastable magnetic phases and the A-AFM ground state confirm that cAFM and FM phases can emerge through thermally-activated spin-flop and spin-flip transitions, respectively, upon increasing the temperature up to the $T_{N}$.

\subsection{Photoluminescence of CrPS$_{4}$ layers $vs$ temperature and magnetic field}
CrPS$_4$ photoluminescence is measured on thin, mechanically exfoliated flakes with thicknesses ranging from 12~nm to 105~nm kept at temperatures between 5~K and 40~K. A magnetic field of up to 10~T is applied either perpendicular or parallel to the surface of the CrPS$_{4}$ flake. The PL signal is excited with a 532~nm continuous wave laser beam and collected through a high numerical aperture lens, in a direction perpendicular to the flake. All studied flakes exhibit quantitatively comparable magneto-optical properties, except for a lower overall PL intensity for thinner flakes.

\begin{figure}[t!]
  \includegraphics[width=0.5\textwidth]{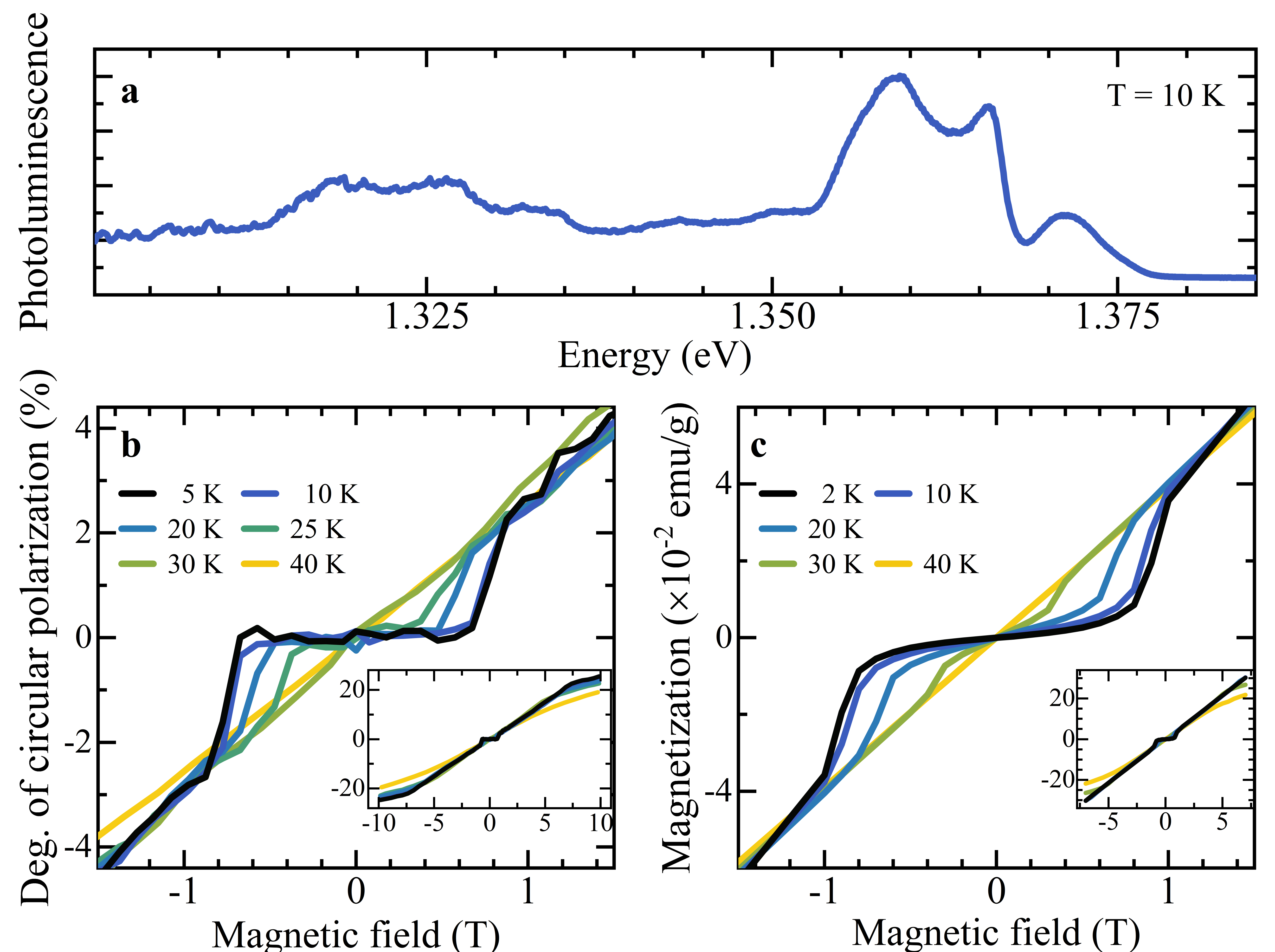}
	\caption{(a) Photoluminescence spectrum of a 105~nm thick CrPS$_{4}$ flake at B = 0~T at the temperature of 10~K. (b) The degree of circular polarization of photoluminescence for the flake in (a), measured in the spectral range \mbox{$1.353~\text{eV} \leq \hbar\omega \leq 1.374$~eV} for temperatures between between 5~K and 40~K. (c) Magnetization curves obtained from SQUID magnetometry for bulk CrPS$_{4}$ for temperatures between 2~K and 40~K. A direct correspondence between the degree of circular polarization and SQUID magnetization is evidenced.}
	\label{fig:magnetoPL_vs_SQUID}
\end{figure}

A PL spectrum of a 105~nm thick CrPS$_{4}$ flake is shown in Fig.~\ref{fig:magnetoPL_vs_SQUID}a. Three distinct maxima are seen in the 1.350~eV - 1.380~eV range. Three weaker maxima showing a comparable intensity pattern as the main ones appear in the range 1.310~eV - 1.340~eV. The emission is attributed to internal $^{4}A_{2g} \rightarrow\, ^{4}T_{2g}$ transitions within the $d$ shell of the Cr$^{3+}$ ion.\cite{Lee:ACSNano2017} %\cite{Susilo:npjQMater2020,Kim:ACSNano2022}
At the same time, electrons from the $d$ shell give rise to the non-vanishing moment of the Cr$^{3+}$ ions responsible for the magnetization of the material. Thus, we anticipate that the observed emission will carry the information on the magnetic state of the crystal. In particular, we expect that the spin projection of the emitted photons on the propagation direction (perpendicular to the CrPS$_{4}$ flake) will reflect the magnetization component of the material parallel to that direction.  In order to verify this hypothesis, we investigate the circular polarization of the PL. 

The degree of circular polarization of the signal is defined as $DOP = (I_{\sigma+}-I_{\sigma-})/(I_{\sigma+}+I_{\sigma-})$, where $I_{\sigma+}$ and $I_{\sigma-}$ are the intensity of the signal registered in $\sigma+$ or $\sigma-$ circular polarizations, respectively. In Fig.~\ref{fig:magnetoPL_vs_SQUID}b the $DOP$ is plotted as a function of the magnetic field applied perpendicular to the CrPS$_4$ flake for consecutive temperatures. Fig.~\ref{fig:magnetoPL_vs_SQUID}c shows corresponding plots of magnetization of bulk material determined by SQUID measurements. A direct correspondence between $DOP$ and magnetization obtained in SQUID experiment is observed. In particular, for temperatures below 10~K, both the DOP and magnetization are negligibly small for magnetic fields below 0.8~T, as expected for the antiferromagnetic CrPS$_4$. Spin-flop type transition of Cr$^{3+}$ ions from A-AFM to cAFM phase is evidenced as an abrupt increase of both the $DOP$ and the magnetization at around $0.8$~T. The increase in  temperature shifts the spin-flop transition towards lower fields. The inset to Fig.~\ref{fig:magnetoPL_vs_SQUID}b shows that the $DOP$ saturates at around 8~T consistent with the tendency seen in the magnetization shown in the inset to Fig.~\ref{fig:magnetoPL_vs_SQUID}c. 

\begin{figure*}[ht!]
\centering
\includegraphics[width=0.9\linewidth]{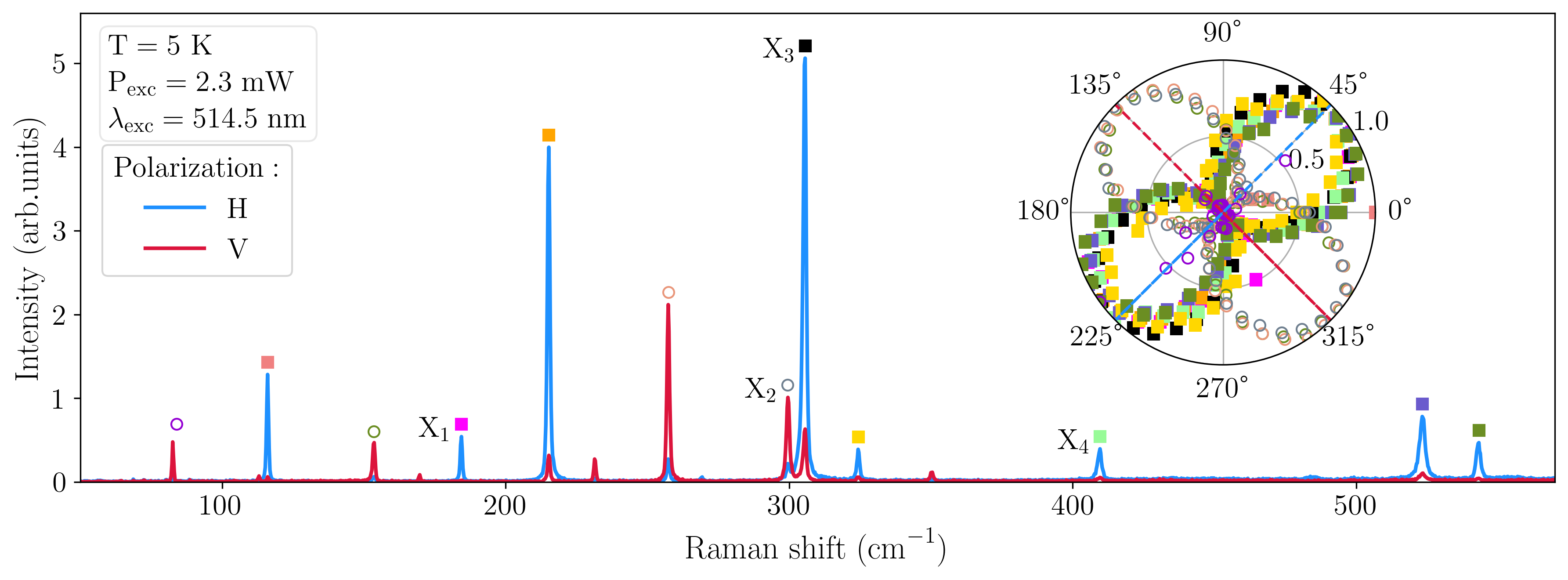}
\caption{Raman spectra of CrPS$_4$ registered in two orthogonal linear polarizations of the detection at T = 5~K for circularly polarized excitation. Phonon modes with $A$ or $B$ symmetry representation are labeled with a full square or an empty dot, respectively. Inset: polar plot of normalized intensity of Raman transitions against the detected linear polarization angle. Despite being excited by a circularly polarized beam, the Raman signal is strongly linearly polarized, with clearly distinguished two perpendicular directions of the polarization, labeled as $H$ and $V$. %$\sigma+$
}
\label{fig:anizo}
\end{figure*} 

A direct proportionality between the $DOP$ and SQUID magnetization is observed also for the magnetic field applied in the plane of the CrPS$_{4}$ flake (see Suppl.~Fig.~\ref{supplfig:PLVoigt}), albeit with a much smaller magnitude of the $DOP$ signal. As discussed above, in our experimental configuration the $DOP$ is primarily sensitive to the out-of-plane magnetization component. However, the use of a high numerical aperture lens to collect the PL signal means that photons emitted at some angle to the normal are also collected, making the measurement also sensitive to the in-plane magnetization to some extent. Additionally, any out-of-plane tilt of the Cr$^{3+}$ spins gives rise to the non-zero $DOP$ signal, even if the spins are mostly aligned with the in-plane magnetic field.

%When the magnetic field is applied in a direction lying in-plane of the flake, the magnitude of the $DOP$ is still measurable, but lower for an order of magnitude as compared to the out-of-plane configuration. Also in this case, however, a direct proportionality between the $DOP$ and SQUID magnetization is evidenced (see Suppl.~Fig.~\ref{supplfig:PLVoigt}).

The critical fields of spin-flop and spin-flip transitions for consecutive temperatures determined from the PL data are superimposed to the magnetic phase diagram shown in Fig.~\ref{fig:magn_phase_diagram}a. A perfect agreement between the magnetospectroscopy and magnetometry data indicates that magneto-photoluminescence provides a very efficient tool in determining the magnetization and magnetic phase diagram of thin layers of CrPS$_4$. Such a direct optical probing of the magnetic properties of this antiferromagnetic material confirms that $d$-shell carriers defining the spin order of Cr$^{3+}$ ions indeed dictate the properties of the optical emission process.

\subsection{Raman spectroscopy of CrPS$_{4}$ layers $vs$ temperature in the absence of magnetic field}

The vibrational spectra of CrPS$_{4}$ yield 36 phonon modes, following symmetry representation $\Gamma_{acoustic} = A + 2B$ and $\Gamma_{optical}= 16A + 17B$. According to the selection rules for the non-centrosymmetric space group \textit{C2} (No.~5), modes with the irreducible representation $A$ and $B$ are Raman-active.\cite{Neal:2DMater2021}

Raman scattering experiments are conducted on bulk CrPS$_{4}$ samples placed in temperatures between 5~K and 250~K. The signal is excited by a circularly polarized laser beam propagating along the crystal's $c$-axis and registered as a function of a detected linear polarization angle. Fig.~\ref{fig:anizo} shows selected spectra acquired for two orthogonal linear polarizations denoted as $H$ and $V$, evidencing well-resolved transitions that represent Raman active modes. Despite the circular polarization of the excitation, the Raman signal is strongly linearly polarized. We attribute this observation to the monoclinic crystal structure of CrPS$_{4}$. 

We extract parameters of the transitions in the Raman spectrum by fitting each line with a Lorentz profile. Experimental transition energies, as well as those calculated within the DFT+U for A-AFM phase, are collected in Suppl.~Table~\ref{Suppltab:phonons}. Their mutually good agreement allows us to assign the calculated modes to the experimental ones and to determine their symmetry.
A polar plot (see the inset to Fig.~\ref{fig:anizo}) shows normalized intensities of the most prominent Raman lines as a function of the detected linear polarization angle. We find that there are two groups of orthogonally polarized lines. The lines which exhibit the same polarization direction have the same representation of their phonon mode, the symmetric $A$ in the case of $H$-polarized lines and anti-symmetric $B$ in the case of $V$-polarized lines.%\cite{Lee:ACSNano2017, Kim:NanoLett2019, Wu:RSCAdv2019, Kim:JPCh2021}

\begin{figure*}[ht!]
\includegraphics[width=0.9\linewidth]{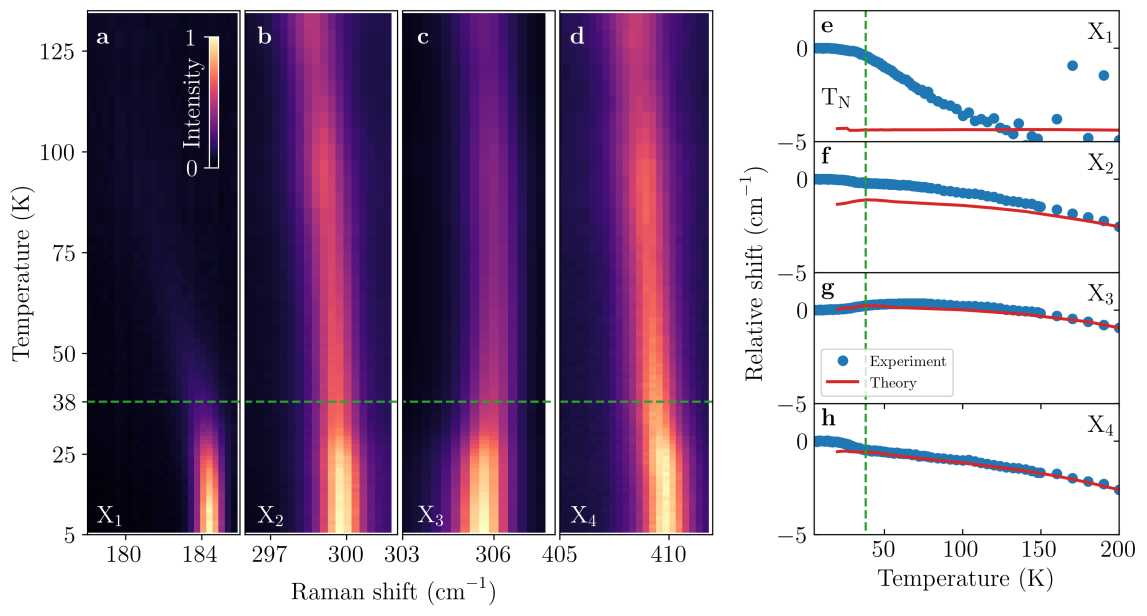}
	\caption{Evolution of (a) X$_1$, (b) X$_2$, (c) X$_3$, (d) X$_4$ transition in Raman spectrum of CrPS$_4$ with temperature. The intensity in panels (a-d) is normalized to the intensity of a given transition at 5~K. (e-g) Position of the lines X$_1$, X$_2$, X$_3$, X$_4$, respectively, as a function of temperature. Full dots: experiment. Lines: a result of the DFT calculations taking as input CrPS$_4$ lattice parameters determined in temperature-dependent XRD measurement. The vertical dashed line marks the N\'{e}el temperature ( 38~K).}
	\label{fig:Raman_spec_vsT}
\end{figure*}

The temperature evolution of selected transitions in the Raman spectrum is shown in Fig.~\ref{fig:Raman_spec_vsT}a-d. Their position, intensity, and linewidth vary with temperature with a qualitative change in the vicinity of the $T_N$, thus reflecting their susceptibility to the type of magnetic ordering in the crystal. The most accentuated change is observed for line X$_1$, positioned at 184.5~cm$^{-1}$ at 5~K (see Fig.~\ref{fig:Raman_spec_vsT}a). Upon approaching the N\'{e}el temperature, the X$_1$ transition weakens and shifts towards lower energies. Above the $T_N$, the shift gets more pronounced, the transition strongly broadens spectrally and finally vanishes around 150~K. Qualitatively similar, but less pronounced redshifts are observed for the lines X$_2$ (at 299.5~cm$^{-1}$, see Fig.~\ref{fig:Raman_spec_vsT}b) and X$_4$ (at 409.8~cm$^{-1}$, see Fig.~\ref{fig:Raman_spec_vsT}d). 

On the contrary, the X$_3$ line at 305.7~cm$^{-1}$ at 5~K shown in Fig.~\ref{fig:Raman_spec_vsT}c shifts toward larger energies (blueshifts) upon approaching and crossing T$_N$. This behavior is qualitatively different with respect to that of lines X$_1$, X$_2$ and X$_4$, which redshift upon approaching the T$_N$ (see Fig.~\ref{fig:Raman_spec_vsT}e-h).
A similar, anomalous, temperature dependence as for X$_3$ transition has been reported previously in the case of selected phonon modes in MnPS$_3$.\cite{Vaclavkova:2DMater2020} It has been attributed to a temperature-induced modification of the superexchange paths between magnetic ions and a resulting change in constants of magnetoelastic interactions between them.\cite{Vaclavkova:2DMater2020} Suitably, our theoretical predictions indicate that X$_3$ shifts towards higher energies when the magnetic configuration departs from the ground, A-AFM state (see Fig.~\ref{fig:modes_diff}).
An evolution of the selected Raman transitions in the temperature range of up to 250~K is presented additionally in Suppl.~Fig.~\ref{fig:Suppl.Ramanspectra_vs_T}.

Suppl. Video SV1 shows an atomic vibrational composition of X$_1$, X$_2$, X$_3$, and X$_4$ Raman lines determined by DFT calculations. In the case of  lines X$_1$ and X$_3$, a predominant contribution comes from in-plane oscillations with an opposite phase of Cr atoms along the $b$ crystal axis. In the case of the X$_2$ line, the CrPS$_4$ crystal cell undergoes torque-like oscillations with the Cr atoms shifting in-plane, in a direction parallel to the $a$ axis of the crystal, yielding a relatively small variation of a pathway length between the Cr atoms. The contribution of the Cr sub-lattice oscillations is even smaller in the case of the X$_4$ line. Hence, in the case of X$_2$ and X$_4$ lines, we attribute the observed spin-phonon coupling to a Cr-S-Cr super-exchange pathway with pronounced oscillations of the S atom placed in between the nearest Cr atoms.

In order to understand the effect of the structural and magnetic phase changes on the Raman modes, we perform the DFT+U calculations with lattice parameters found in the temperature-dependent X-ray diffraction measurement (see Fig.~\ref{fig:XRD}). Previous DFT calculations of the Raman modes in 2D magnetic materials assumed optimized lattice parameters and neglected their temperature dependence. In this regard, the structural changes determined by XRD are accounted for by the respective rescaling of the lengths of the optimized cells (for the details, see Sec.~\ref{SuplSec:modelling} in Suppl. Information). 

As shown in Fig.~\ref{fig:Raman_spec_vsT}e-g, the calculated temperature evolution for most Raman modes in the case of the most of the modes generally follows the experimental trends (see also Suppl.~Fig.~\ref{Supplfig:teorshiftvsT}). The calculated and experimental changes are of the same order of a few cm$^{-1}$ in the studied temperature range, and the slopes of the dependencies increase with the temperature. As the calculated evolution reflects changes in the lattice parameters, the observed temperature-induced shifts of the presented Raman modes above the T$_N$ are primarily of structural origin. On the other hand, slopes of the calculated and experimental dependencies qualitatively disagree below T$_N$ for the lines X$_2$ and X$_4$. Also, the dependence determined theoretically for the X$_1$ line does not follow the experimental trend, but shows a surprising insensitivity to temperature-induced structural changes. In order to give a physical insight into a plausible mechanism of the observed discrepancies, the impact of various magnetic arrangements of the Cr$^{3+}$ spins on the Raman modes is examined in the next Section.

\begin{figure}[t!]
  \includegraphics[width=0.5\textwidth]{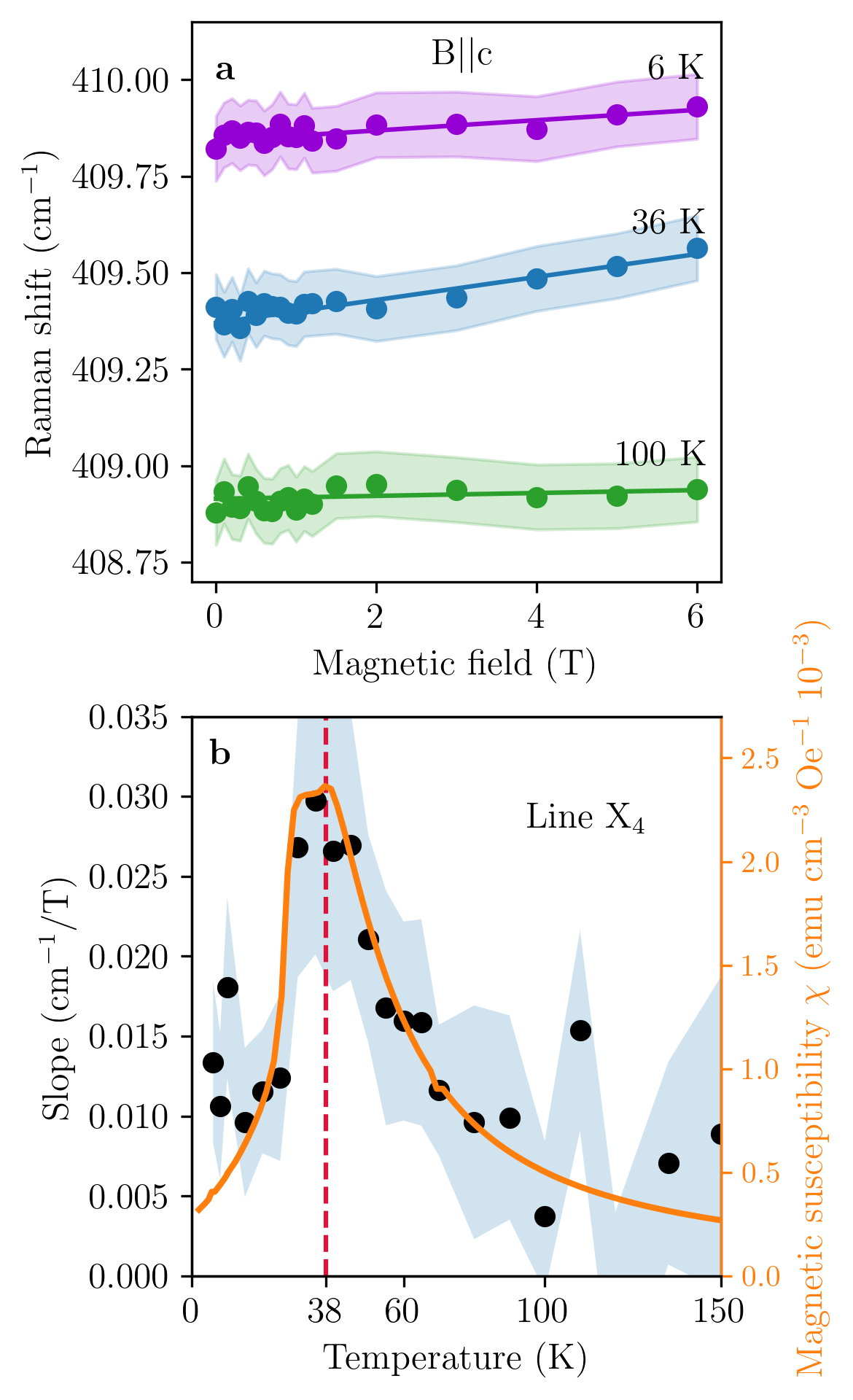}
	\caption{(a) Spectral position of the X$_4$ transition in Raman spectrum of CrPS$_{4}$ as a function magnetic field applied along the $c$ crystal axis for temperatures 6~K, 35~K and 100~K (points). Lines represent linear fit to the experimental data. (b) Slopes of the linear dependencies found for consecutive temperatures in the range from 5~K to 150~K (points) along with the magnetic susceptibility curve obtained from SQUID magnetometry (orange line) at B = 0.05~T. Shaded regions in both panels encodes uncertainty of the Raman data. The magnitude of the magnetic-field-induced shift of the X$_4$ line follows the magnetic susceptibility curve from magnetometry.} %in $\sigma^-$ polarization
	\label{fig:Raman_vs_suscept}
\end{figure}

\subsection{Raman spectroscopy of CrPS$_{4}$ layers $vs$ magnetic field and temperature}
%unpolarized excitation - a laser He-Cd, detection - circular
%The detected polarizations are $\sigma^{+}$ and $\sigma^{-}$.
Magneto-Raman measurements are performed in the Faraday configuration, with the magnetic field applied parallel to the $c$-axis of the crystal. Although in ordinary semiconductors, magnetic field influences phonon frequencies only weakly, in the case of the studied material incorporating magnetic Cr$^{3+}$ ions in its lattice, the impact of the magnetic field on Raman spectra is sizeable. 

Fig.~\ref{fig:Raman_vs_suscept}a shows the spectral position of the X$_4$ line (points) as a function of the magnetic field for three selected temperatures: 6~K, 35~K and 100~K. The X$_4$ line shifts towards higher energies with the increasing field, following an approximately linear dependence. The linear fits (solid lines in Fig.~\ref{fig:Raman_vs_suscept}a) performed for the data registered at temperatures in the range between 6~K and 150~K yield slopes, which are represented by black points in Fig.~\ref{fig:Raman_vs_suscept}b. The magnetic susceptibility curve obtained from the magnetometry measurements (see Suppl.~Fig.~\ref{fig:sus}a) at a small field of 0.05~T in the parallel configuration is also displayed in Fig.~\ref{fig:Raman_vs_suscept}b with an orange line. As clearly seen, the slopes of the magnetic-field-induced energy shift of the X$_4$ line follow the magnetic susceptibility curve of the material. In particular, the maxima of both dependencies in Fig.~\ref{fig:Raman_vs_suscept}b overlap, which means that the susceptibility of the phonon mode X$_4$ energy to the magnetic field is the largest in the vicinity of the N\'{e}el temperature. Moreover, the magnetic-field-induced change in the degree of circular polarization of the X$_4$ line is strongly reduced above the N\'{e}el temperature (see Suppl.~Fig.~\ref{supplfig:Ramanpoldeg}), which marks the magnetic phase transition taking place at T$_N$. 

The results described above demonstrate that, by performing temperature and magnetic-field-dependent Raman scattering measurement, one can determine the shape of the magnetic susceptibility curve and state the presence of magnetic phase transitions of 2D layered antiferromagnet CrPS$_4$. The theoretical calculations presented below indicate which transitions in the Raman spectrum of CrPS$_4$ are the best suited for this purpose. 

\begin{figure}[h!]
\centering
\includegraphics[width=0.45\textwidth]{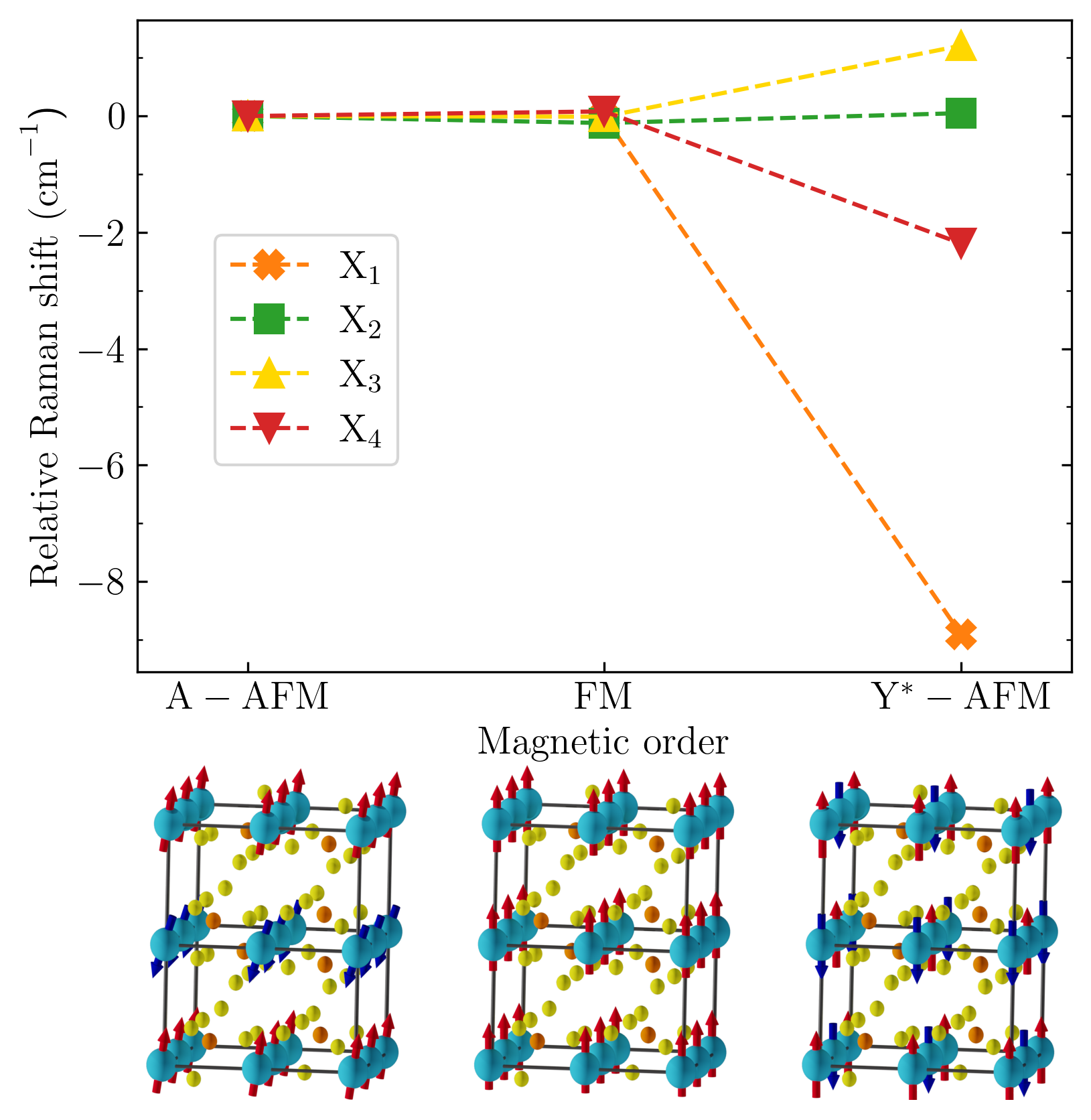}
\caption{Calculated relative spectral shift of the X$_1$, X$_2$, X$_3$ and X$_4$ Raman modes upon a transition from the ground-state A-AFM to out-of-plane-spin-flipped FM and in-plane-spin-flipped Y*-AFM magnetic phase of CrPS$_{4}$. A schematic representation of the spin order of Cr$^{3+}$ ions in each magnetic phase is shown below the horizontal axis of the plot.}
\label{fig:modes_diff}
\end{figure}

We consider the impact of the in-plane and out-of-plane arrangements of the Cr$^{3+}$ ions spins on optically active Raman modes, examining the lowermost energetically, collinear magnetic phases, \emph{i.e.} A-AFM, FM and Y$^*$-AFM. Ferromagnetic order within the $a-b$ crystal plane is maintained in AFM and FM phases. The Y$^*$-AFM is a prototype phase, where spins of the Cr$^{3+}$ first nearest neighbors (1NN) are antiferromagnetically aligned, leading to a suppression of the the FM ordering within the $a-b$ plane. In view of our DFT calculations, the change of the type of out-of-plane magnetic ordering (e.g., A-AFM to FM) shifts Raman modes only by fractions of cm$^{-1}$ in relation to unperturbed modes (see Fig.~\ref{fig:modes_diff} and Suppl.~Table~\ref{Suppltab:phonons}). However, we predict pronounced changes in the spectral position of the Raman modes (of the order of cm$^{-1}$) when a magnetic phase transition that alters the in-plane spin ordering occurs (see Fig.~\ref{fig:modes_diff} and Suppl. Table~\ref{Suppltab:phonons}). In particular, a transition from A-AFM to Y$^*$-AFM results in a 9~cm$^{-1}$ redshift in the case of the X$_1$ line, thus confirming its strong susceptibility to variation of the magnetic order revealed in the experiment (see Fig.~\ref{fig:Raman_spec_vsT}a). Our predictions are in line with the calculated strength of the exchange coupling (see Suppl.~Table~\ref{Suppltable:exchconst}), which is order of magnitude stronger (J$_1 = 2.13$~meV) for in-plane than for out-of-plane (J$_0 = -0.14$~meV) directions. Calculations confirm also the enhanced proximity between next-neighboring Cr atoms along the $b$-direction in the crystal. These findings are consistent with the results of temperature-dependent XRD studies (Fig.~\ref{fig:XRD} and Suppl.~Fig.~\ref{fig:XRD_suplement}), showing that the $b$ lattice parameter of the crystal diminishes with the emperature. This leads to an increased probability of magnetic states other than ground A-AFM in the material (see Suppl.~Tab.~\ref{Suppltab:magorders_energy}). 

%Further theoretical considerations show that 1NN and second nearest neighbour (2NN) distances between the Cr atoms diminish upon a transition from the ground AFM phase to a phase with in-plane AFM order (see Suppl.~Fig.~\ref{Supplfig:dist_vs_order}). 

As Fig.~\ref{fig:modes_diff} reveals, a relatively strong redshift assisting the AFM to Y$^*$-AFM transition is predicted in the case of the X$_4$ line. Since this line, in contrast to the line X$_1$, is still observable above the $T_N$, we have selected it for the magneto-optical Raman study presented in Fig.~\ref{fig:Raman_vs_suscept} and Suppl.~Fig.~\ref{supplfig:Ramanpoldeg}. A sizeable spectral shift predicted for X$_3$ (see Fig.~\ref{fig:modes_diff}) suggests that the magnetic susceptibility curve can be retrieved also from this line's magnetic field and temperature spectral dependencies. Suppl.~Fig.~\ref{fig:Raman_vs_suscepX3} shows that this is indeed the case. We note, however, that magnetic field-induced spectral shifts of X$_3$ provide an inverted temperature dependence with respect to that observed for the X$_4$ line (compare Suppl.~Fig.~\ref{fig:Raman_vs_suscepX3} for the X$_3$ line with Fig.~\ref{fig:Raman_vs_suscept} for the X$_4$ line). This observation is consistent with the theoretical considerations of Fig.~\ref{fig:modes_diff}, for which opposite spectral shifts are expected for X$_3$ (blueshift) and X$_4$ (redshift) at the AFM to Y$^*$-AFM transition. In turn, the X$_2$ line shows no susceptibility to a magnetic order change, as can be seen in Fig.~\ref{fig:modes_diff}. Accordingly, a magneto-optical Raman study of the X$_2$ line does not bring any meaningful dependencies (not shown). 

The consistency between the presented set of magneto-optical and DFT results 
shows that the calculations enable the identification of phonon modes in the Raman spectrum, which are best suited for determination of a susceptibility curve of the material. %The calculations predict also a qualitative character of the dependencies observed in temperature dependent Raman magnetospectroscopy.

\section{Conclusions}

We have performed a multidimensional study on the layered antiferromagnet CrPS$_4$ involving magnetometry, structural and magneto-optical measurements and complementary DFT theoretical calculations. 

We find that the degree of circular polarization of photoluminescence of CrPS$_4$ mirrors magnetization determined in SQUID measurements. This direct correspondence can be traced back to the $d$-shell electrons responsible for both the emission and magnetization of the material. Moreover, Raman scattering measured as a function of magnetic field and temperature enables the determination of the magnetic susceptibility curve of the material. The temperature evolution of the optically active Raman modes above the N\'{e}el temperature agrees well with theoretical predictions taking into account structural changes of the crystal determined in temperature-dependent XRD measurements. The discrepancies observed at temperatures at or below $T_N$ are a fingerprint of the susceptibility of the Raman modes to the magnetic arrangement of the Cr$^{3+}$ ions. The calculations confirm a strong magnetoelastic coupling in CrPS$_4$ and allow us to explain anomalous dependencies in the thermal expansion of the crystal lattice and temperature shifts of selected Raman lines. The theoretical considerations allow us also to predict the influence of the type of spin arrangement of Cr$^{3+}$ ions on the Raman modes in CrPS$_4$, and to identify Raman modes of the material that are the most susceptible to the applied magnetic field. 

With our work, we show that optical magnetospectroscopy, \emph{i.e.}, a purely optical measurement, is capable of yielding magnetometry-type information on the properties of an antiferromagnetic semiconductor, including its magnetization or the magnetic phase diagram. We envision that the presented approach will be useful in the case of thin layers of other 2D or ordinary-type magnetic materials, for which standard methods for the determination of magnetic susceptibility curves do not provide a sufficient signal-to-noise ratio.

% revealing that the Raman frequency shifts of lines above the N\'{e}el temperature are of mainly of structural origin.
%thus those, which are optimal for the purpose of determination of the magnetic susceptibility curve of the material. 

\section{Methods}

\subsection{Synthesis of CrPS$_{4}$ samples}

Bulk CrPS$_{4}$ samples are synthesized by the physical vapor transport - without any transporting agent such as iodine (see Ref.\cite{Budniak:Small2020} for details). Around 1 gram of stoichiometric mixture of elements, with the ratio Cr : P : S = 1 : 1 : 4 is sealed in a quartz ampule previously evacuated by a turbomolecular pump. The ampule is kept for 4 days in a furnace with a temperature gradient of 750°C/710°C (substrate zone/deposition zone) and after cooling down a few mm by a few mm CrPS$_{4}$ crystals with a thickness of 10-500 $\mu$ are collected.

\subsection{XRD Investigation}

X-ray diffraction measurements are conducted on CrPS$_{4}$ samples of two types: single crystals and powdered crystals. Diffractograms of single crystals are acquired on samples placed in the preferential orientation (001) in a parallel beam mode by Rigaku SmartLab 9~kW, Cu K$\alpha$ radiation 1.5406 Å, equipped with a Ge 220 monochromator.

Temperature-dependent non-ambient X-ray diffraction measurements on powdered samples are performed using MalvernPanalytical Empyrean diffractometer equipped with Oxford Instruments PheniX helium close-cycle refrigerator. Measurements are  made in 20 - 300~K temperature range with the step of 2~K (in the range $20 - 60$~K) or 20~K ($60 - 300$~K). Thermal displacement of the sample stage is automatically corrected by reference measurements of a high purity tungsten (6N) powder. During the measurements high-vacuum conditions (10$^{-7}$~mbar) are maintained. Collected patterns are refined using FullProf Suite package.

\subsection{Electron Microscopy}

HR-SEM Zeiss Ultra-Plus FEG-SEM working on 4~keV is used to acquire a picture with a secondary electron detector. EDS spectrum is recorded by FEI E-SEM Quanta 200 with 20 kV acceleration voltage. In both cases, the sample is not coated.

CrPS$_{4}$ is mechanically exfoliated directly onto a TEM grid by the protocol developed in Ref.\cite{Budniak:Small2020} In the present work, the use of the tape: “Ultron systems, INC. Silicon-Free Blue Adhesive Film P/N 1009R-6.0“, increases the entire protocol efficiency compared to the previous works.\cite{Budniak:AdvOptMat2022}

In the first step, the mechanically exfoliated CrPS$_{4}$ is measured by cTEM: FEI Tecnai G$^{2}$ T20 S-Twin TEM working on 200 keV. The average background subtraction filter (ABSF) is applied for the micrograph presented in the Suppl.~Fig. ~\ref{fig:SI_structural_characterization}d. Then, the previously prepared TEM grid with CrPS$_{4}$ is mapped by high angle annular dark-field (HAADF) and EDS detectors in STEM mode with 60 kV acceleration voltage on FEI Titan Cubed Themis G$^{2}$ 60–300 equipped with a Dual-X detector (Bruker) with an effective solid angle of 1.76 sr for fast and precise local atomic chemical analysis.

The micrograph presented in Fig.~\ref{fig:HRSTEM} is based on 20 averaged and aligned frames acquired by HAADF detector. EDS maps in Fig.~\ref{fig:HRSTEM} are based on 200 frames. All EDS maps are pre-filtrated by pixel averaging (7 px) and post-filtrated by radial Wiener filtering (highest frequency 15.0 and edge smoothing 5.0) with Velox Software (Thermo Fisher Scientific). Low voltage (60~kV) for electron acceleration is applied to minimize damage of the material.

\subsection{Magnetometry measurements}

Magnetometry measurements are performed using a SQUID magnetometer on a CrPS$_{4}$ bulk crystal, with approximate dimensions 3~mm~$\times$~2~mm~$\times$~0.5~mm and density of 2.64~g/cm$^3$. The sample dipolar magnetic moment is measured along the applied magnetic field $B$. Measurements are performed in two configurations: in a parallel one, with $B \parallel$ to the $c$-axis of the crystal and in a perpendicular one, with $B \perp c$.

\subsection{Optical magnetospectroscopy}
Photoluminescence from CrPS$_{4}$ layers of a thickness between 12~nm and 105 nm is excited with a beam of 561~nm laser. The Raman spectroscopy measurements on bulk CrPS$_{4}$ samples are performed using a linearly polarized 514.5~nm line of the Ar-ion laser or an unpolarized 442~nm line of the He-Cd laser, at temperature varied from 5~K to 250~K. 

The lens of 3.5~mm focal length mounted above the sample on piezo-translators ensures the spatial resolution of 1.5~$\mu$m. Magnetic field of up to 10~T produced by a superconducting coil is applied in Faraday or Voigt configuration. The excitation and detection is polarization resolved thanks to the use of a relevant combination of a linear polarizer, a half-wave plate with a controlled angle and a quarter-wave plate. The signal is detected using a 0.75~m grating spectrometer (2400~grooves/mm) with a Peltier cooled CCD camera on its output (0.1~nm of spectral resolution).

\subsection{Computational details}

Planewave DFT calculations have been performed using VASP package\cite{KRESSE199615} with PAW pseudopotentials\cite{Holzwarth2001} and the PBE functional\cite{PhysRevLett.77.3865,doi:10.1063/1.1926272}. On-site Coulomb repulsion correction is introduced via the DFT+U approach introduced by Dudarev \textit{et al.}\cite{Dudarev:PRB1998} U~=~3~eV and J~=~1.4~eV (U$_{eff}=1.6$~eV) parameters have been chosen to account for the repulsion between 3d electrons of Cr atoms.\cite{Houlong:PRB2016} We note that a comparable value of U (U = 1.8 eV) selected previously in the case of exfoliated MnPS$_3$ layers has been corroborated by the recent ARPES measurements.\cite{Strasdas2023} Each magnetic phase of CrPS$_4$ is modeled with a 1$\times$1$\times$2 supercell of the conventional cell (8 formula units). The unit cell shape, volume and atomic positions are relaxed until forces acting on each atom are smaller than 0.001~eV/Å. The electron density is optimized until 10$^{-7}$~eV self-consistency criterion is met. A $\Gamma$-centered 8$\times$8$\times$4 k-point grid is used for all calculations.

The effect of temperature is modeled using experimental lattice parameters obtained via refinement of XRD data obtained in the temperature range of 20 -- 300~K. The various magnetic arrangements have been adopted as presented in Suppl.~Fig.~\ref{Supplfig:spinarr}. For each of the employed magnetic orderings the position of the atoms have been fully relaxed. For each magnetic ordering independently, relaxed lattice parameters $a, b, c$ and $\beta$ are scaled at each temperature by such factor as found in XRD experiment. $\Gamma$-point phonon frequencies are calculated using the finite displacement method with a 2$\times$1$\times$2 supercell of the primitive unit cell (8 formula units). The self-consistent convergence criterion is increased to 10$^{-11}$ eV to accurately evaluate ionic forces. The phonopy\cite{Phonopy} package is used to generate displacements, build the dynamical matrix from VASP-calculated forces, and calculate the phonon frequencies.

\section{Acknowledgments}

M. B. acknowledges financial support from the  University of Warsaw under the "Excellence Initiative - Research University" project. J. S. and M. G. acknowledge support within the "New Ideas 2B in POB II” IDUB project financed by the University of Warsaw. We acknowledge Polish high-performance computing infrastructure PLGrid for awarding this project access to the LUMI supercomputer, owned by the EuroHPC Joint Undertaking, hosted by CSC (Finland) and the LUMI consortium through PLL/2022/03/016435. A. K. B. and E. L. were supported by the European Commission via the Marie Skłodowska-Curie action Phonsi (H2020-MSCA-ITN-642656). Research leading to these results has also received funding from the Norwegian Financial Mechanism 2014 to 2021 under Grant No. 2020/37/K/ST3/03656 and from the Polish National Agency for Academic Exchange within Polish Returns program under Grant No. PPN/PPO/2020/1/00030.

\bibliography{main}

 %%%% SUPPLEMENTARY INFORMATION %%%%

%\pagebreak[4]
\clearpage

\begin{abstract}
This file contains supplementary information to the article "Direct Optical Probing of the Magnetic Properties of the Layered Antiferromagnet CrPS$_4$". It includes additional information regarding a structural characterisation of the samples, their magnetic properties, additional results of temperature and magnetic field dependent Photoluminescence and Raman spectroscopy measurements, and description and results of theoretical modelling.
\end{abstract}

\title{Supplementary Information to "Direct Optical Probing of the Magnetic Properties of the Layered Antiferromagnet CrPS$_4$"}

\maketitle

%\newpage
\setcounter{page}{1}
\setcounter{figure}{0}
\setcounter{table}{0}
\setcounter{section}{0}

\renewcommand{\thepage}{S\arabic{page}}
\renewcommand{\thesection}{S\arabic{section}}
\renewcommand{\thetable}{S\arabic{table}}
\renewcommand{\thefigure}{S\arabic{figure}}

\onecolumngrid
\section{\label{SI:struct characterization} Structural characterization}

\begin{figure*}[h!]
\centering
 \includegraphics[scale=0.4]{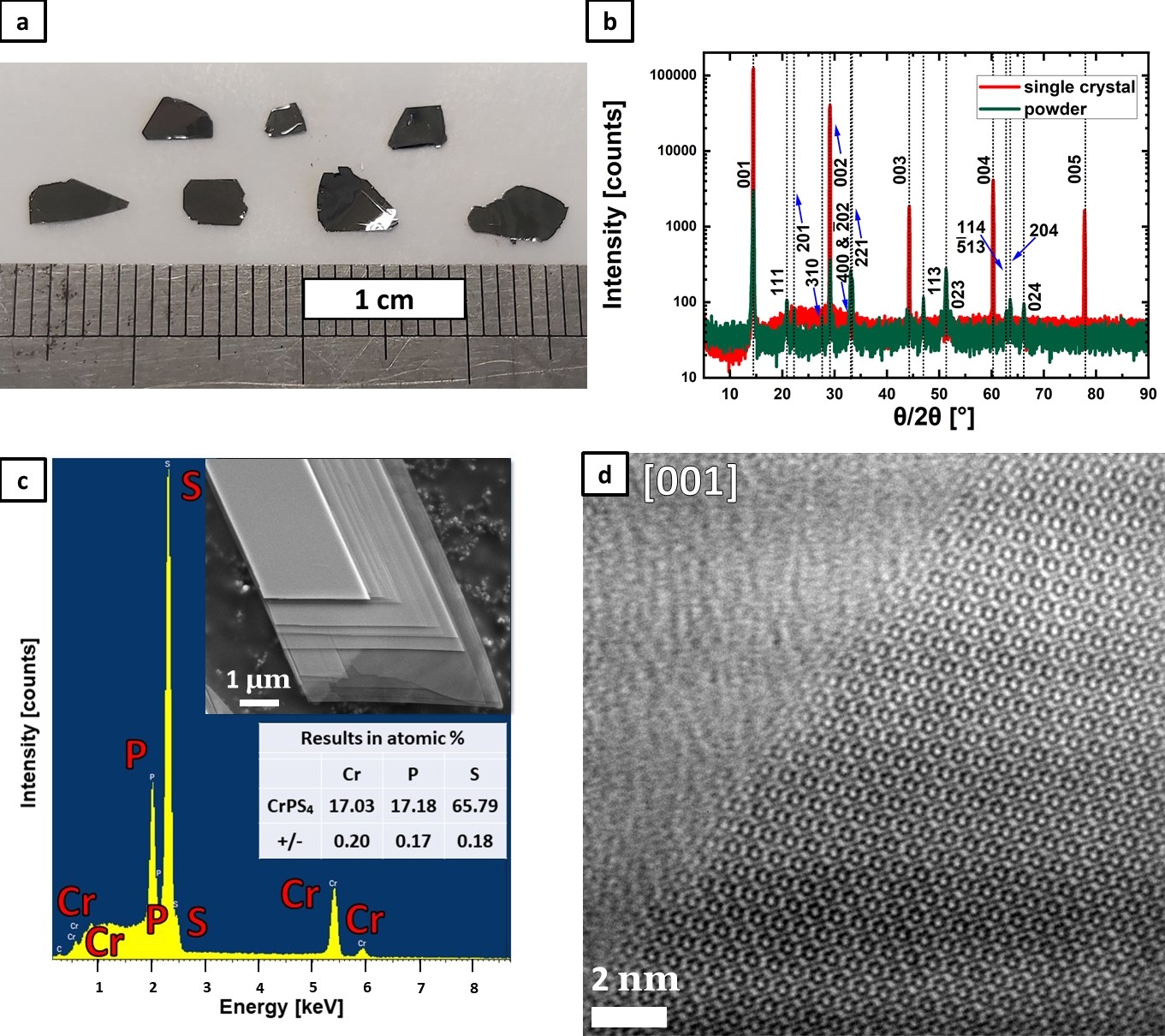}
\caption{(a) An image of "as-synthesized" bulk CrPS$_{4}$ crystals. (b) Results of X-ray studies of CrPS$_{4}$: a diffractogram of a single CrPS$_{4}$ crystal aligned in a preferential orientation that is with the $c$-axis parallel to the X-ray beam propagation (red curve) and of powdered CrPS$_{4}$ crystal (green curve). Vertical black dotted lines describe CrPS$_{4}$ peak positions according to PDF 04-009-0826.\cite{Louisy:SSC1978} (c) EDS spectrum of a bulk CrPS$_{4}$ crystal. The obtained atomic composition shown in the table confirms Cr:P:S atomic ratio close to 1:1:4. The inset: HR-SEM micrograph of a bulk CrPS$_{4}$ sample recorded with a secondary electron detector shows a layered structure of the crystal. (d) A high-resolution, filtered micrograph of a mechanically exfoliated CrPS$_{4}$ layer registered with a conventional TEM.}
\label{fig:SI_structural_characterization}
\end{figure*}

The results of a general structural characterization of bulk CrPS$_{4}$ are presented in Suppl. Fig.~\ref{fig:SI_structural_characterization}. A photograph of "as-synthesized" chromium thiophosphate crystals with a size of the order of half of a centimeter and a thickness of the order of 10 $\mu$m - 100 $\mu$m is shown in Suppl.~Fig.~\ref{fig:SI_structural_characterization}a. One of the synthesized crystals is placed on a flat surface to register a diffractogram using a powder X-ray diffractometer (PXRD), which is shown with a red curve in Suppl. Fig.~\ref{fig:SI_structural_characterization}b. Only a family of 00I planes is registered, which is expected in the case of the analysis conducted on a single crystal of a layered material, in which van der Waals planes are perpendicular to the scattering vector. In contrast, the powder diffractogram presented via the green line reveals all planes present in CrPS$_{4}$. The position of Bragg peaks in both diffractograms remains in agreement with the literature.\cite{Louisy:SSC1978} 

Energy-dispersive X-ray spectrum (EDS) shown in Suppl. Fig.~\ref{fig:SI_structural_characterization}c confirms the presence of all three elements, that are chromium, phosphorus, and sulfur. Results of a quantitative analysis are provided in the table in the same panel. They corroborate that the atomic composition of the studied crystal is close to the intended Cr:P:S = 1:1:4. The inset of Suppl. Fig.~\ref{fig:SI_structural_characterization}c presents a scanning electron microscope (SEM) picture of CrPS$_{4}$, with its clearly evidenced layered structure. In addition, it can be noticed that the crystal cleaves along crystallographic directions.

Suppl.~Fig.~\ref{fig:SI_structural_characterization}d presents a transmission electron microscope (TEM) micrograph of the crystal mechanically exfoliated onto a grid. The image registered in the [001] zone axis confirms a good structural quality of the studied sample. 

\begin{figure*}[h!]
\centering
 \includegraphics[scale=0.5]{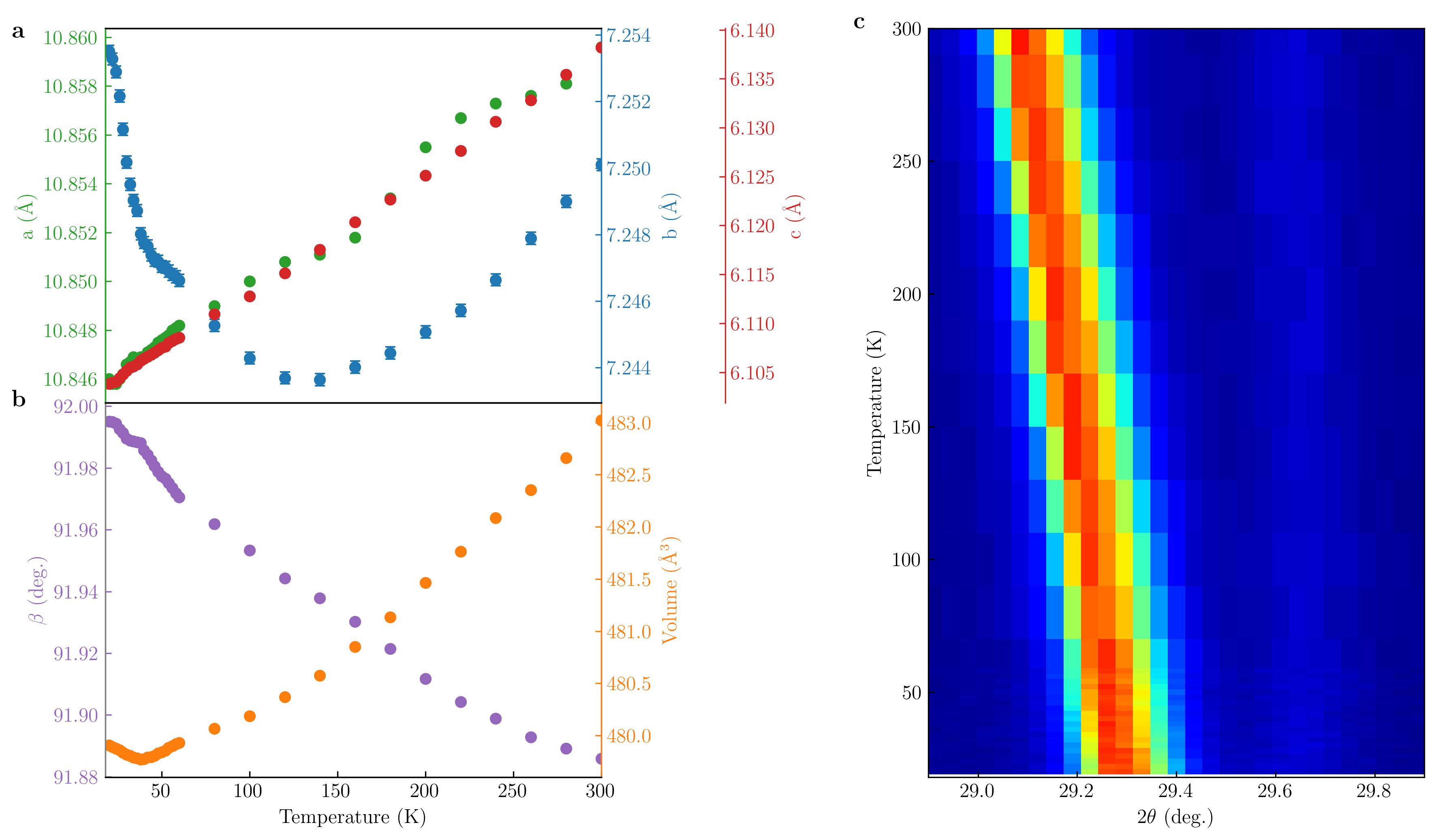}%XRD_supplement.png
\caption{Results of non-ambient XRD studies on powdered samples of CrPS$_{4}$ in the entire investigated temperature range (20~K - 300~K): (a) $a$, $b$ and $c$ lattice parameters (b) $\beta$ angle and unit cell volume as a function of temperature (c) Variation of the 002 and 220 reflections with the temperature.}
\label{fig:XRD_suplement}
\end{figure*}

Suppl. Figure~\ref{fig:XRD_suplement} shows the results of XRD measurements as a function of temperature in the range from 20~K to 300~K on a powdered CrPS$_{4}$. Suppl.~Figures~\ref{fig:XRD_suplement}a and b show the variation of the lattice parameters. The lattice parameters at 300~K are $a$ = 10.8596(1)~\r{A}, $b$ = 7.2501(2)~\r{A}, $c$ = 6.1382(2)~\r{A}, $\beta$ = 91.886(1)$^{\circ}$, and cell volume is 483.02(1)~\r{A}$^{3}$, in agreement with the results of our STEM measurements and previous reports.\cite{Diehl:ActaCryst1977} Suppl.~Fig.~\ref{fig:XRD_suplement}c presents additionally a map composed of consecutive XRD spectra containing selected two maxima (reflections 002 and 220) as a function of temperature in the range between 20~K and 300~K.

\newpage
\newpage
\clearpage

\section{Magnetometry characterisation}

\begin{figure*}[h!]
\centering
\includegraphics[scale=0.75]{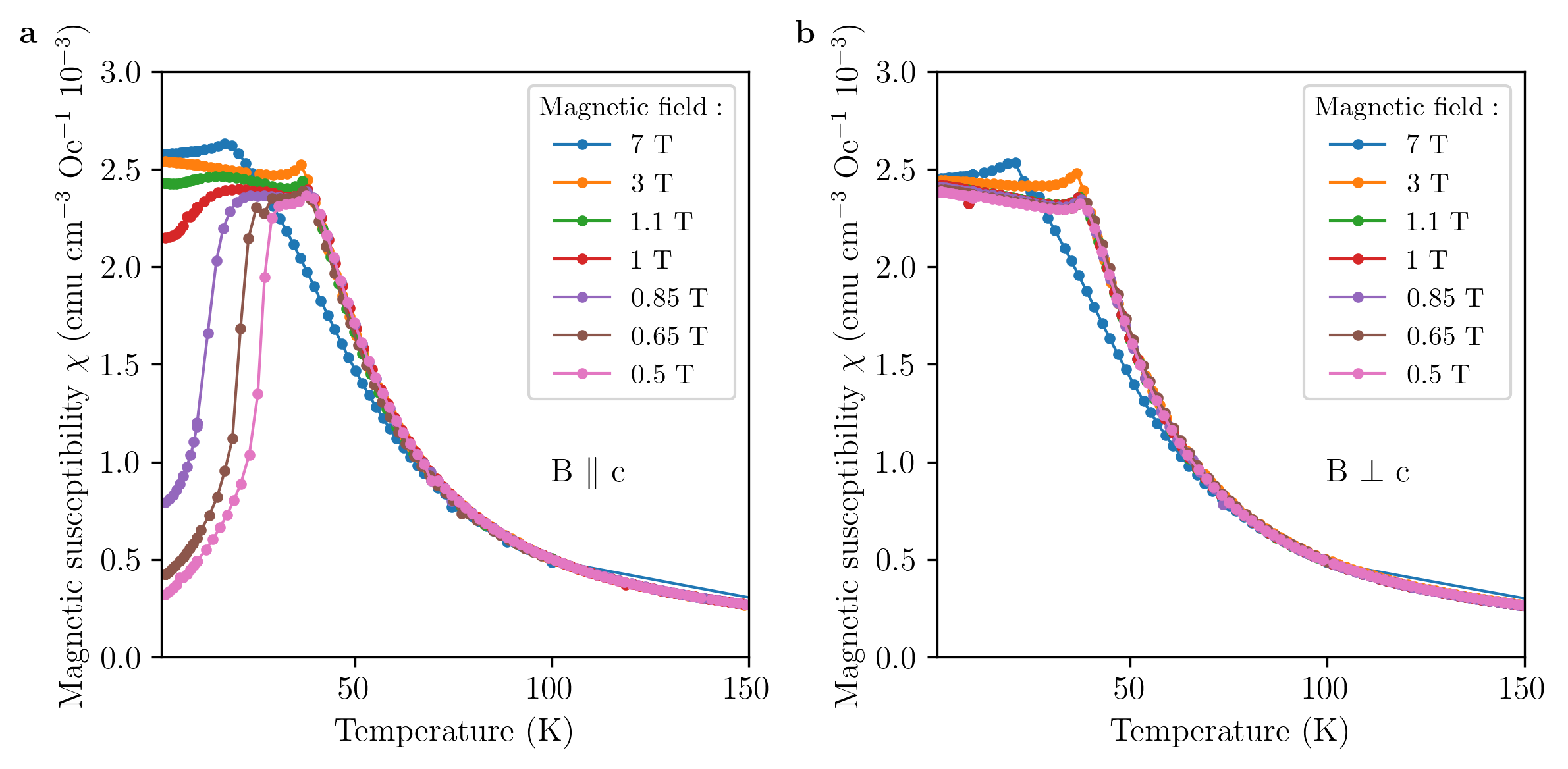}
\caption{Magnetic susceptibility $vs$ temperature of a bulk CrPS$_{4}$ sample for the magnetic field applied in a direction (a) parallel and (b) perpendicular to the crystal $c$-axis. The N\'{e}el temperature marked as the maximum of the curve is 38~K in the absence of magnetic field and shifts noticeably towards lower temperatures for magnetic fields exceeding 1.1~T. An additional maximum below the main peak observed at low field values is attributed to a transition from A-AFM to canted AFM magnetic phase of the crystal.}
\label{fig:sus}
\end{figure*}

Measurements of magnetic characterization of the CrPS$_{4}$ bulk sample correspond to the longitudinal magnetic moment along the applied field. The results acquired with the magnetic field applied in the direction parallel and perpendicular to the $c$-axis of the crystal are presented respectively in Suppl.~Fig.~\ref{fig:sus}a and Suppl.~Fig.~\ref{fig:sus}b. Phase diagram constructed from the data collected in both parallel and perpendicular configurations is shown in Fig.~\ref{fig:magn_phase_diagram} in the main text.

\newpage
\newpage

\section{Mangeto-Photoluminescence}

 \begin{figure*}[h!]
\centering
 \includegraphics[width=0.8\textwidth]{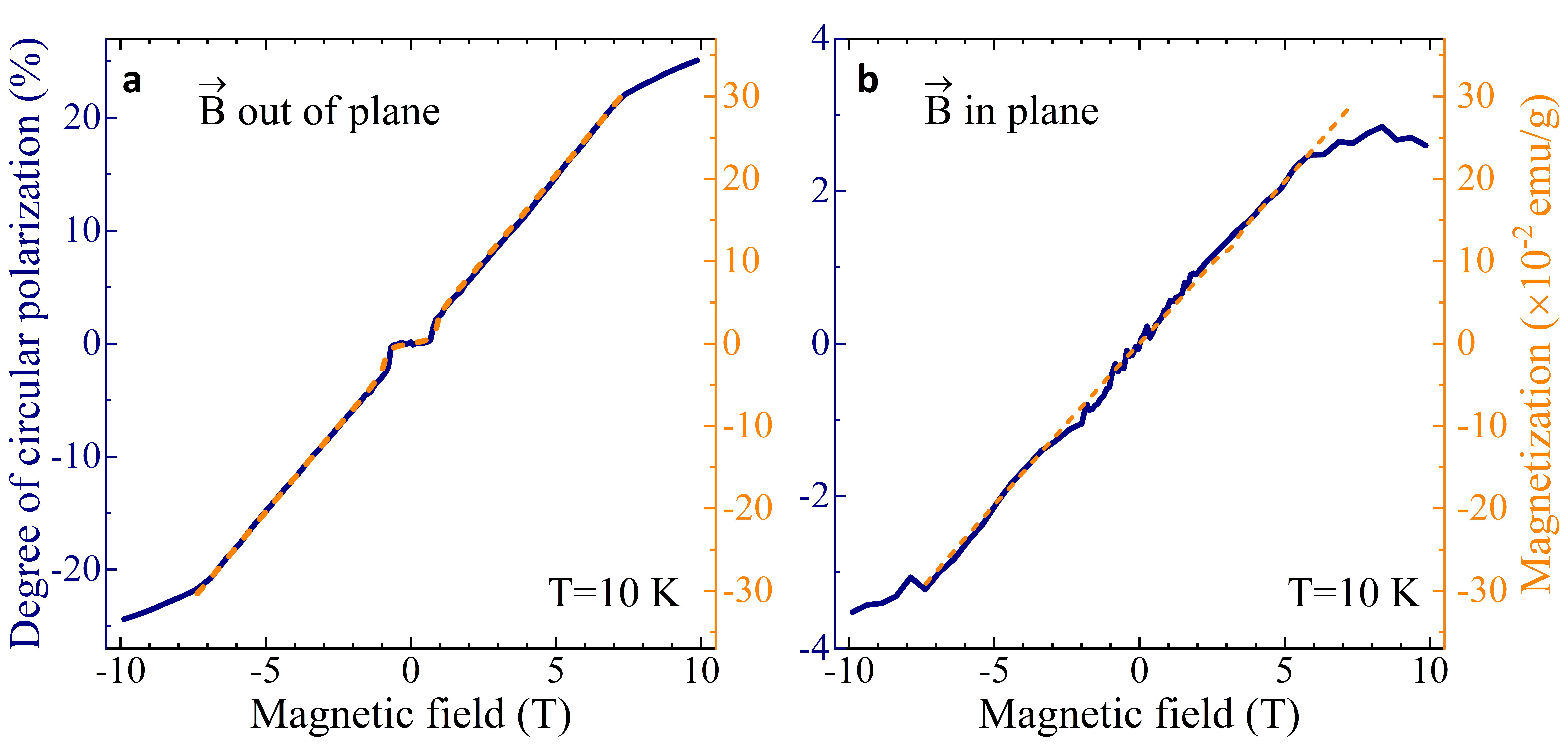}
\caption{The degree of circular polarization ($DOP$, blue curves) of photoluminescence of the 105~nm thick CrPS$_{4}$ flake, determined for the spectral range \mbox{$1.353~\text{eV} \leq \hbar\omega \leq 1.374$~eV}~eV at the temperature of 10~K in two configurations of magnetic field: (a) normal to the crystal surface, and (b) in-plane of the crystal. Light emitted from the sample is detected in the direction perpendicular to the surface of the flake. Orange dashed curves represent the magnetization curves obtained from SQUID magnetometry for bulk CrPS$_{4}$ at the temperature of 10 K for the same configurations of the magnetic field. Note that the $DOP$ signal in panel (b) has a significantly smaller amplitude than in the case of panel (a), as in our experimental configuration it is primarily sensitive to the out-of-plane magnetization component. However, the use of a high numerical aperture lens to collect the PL signal means that photons emitted at some angle to the normal are also collected, making the measurement also sensitive to the in-plane magnetization to some extent. Additionally, any out-of-plane tilt of the Cr$^{3+}$ spins gives rise to the non-zero $DOP$ signal, even if the spins are mostly aligned with the in-plane magnetic field.}
\label{supplfig:PLVoigt}
\end{figure*}

\newpage
\section{Raman spectroscopy}

 \begin{figure*}[h!]
\centering
 \includegraphics[width=0.8\textwidth]{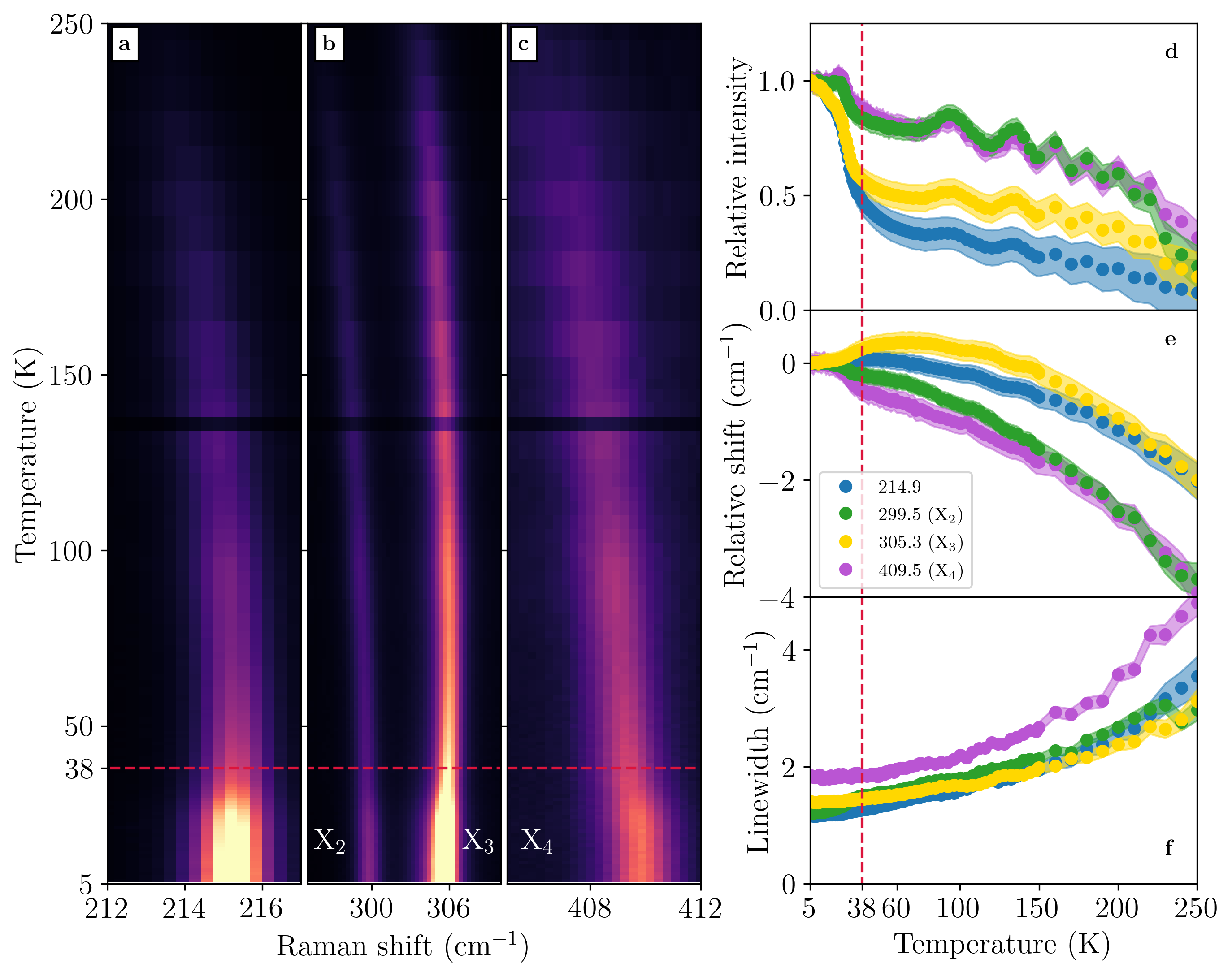}
\caption{Evolution of selected transitions in the Raman scattering spectrum of bulk CrPS$_{4}$ with the increase of temperature from 5~ K to 250~K: (a) line at 214.9~cm$^{-1}$, (b) lines at 299.5~cm$^{-1}$ (X$_2$ in the main text) and 305.7~cm$^{-1}$ (X$_3$ in the main text), (c) line at 409.8~cm$^{-1}$ (X$_4$ in the main text). (d) The amplitude, (e) spectral shift, and (f) linewidth of the transitions shown in panels a-c as a function of temperature. The normalized intensities and spectral shifts are relative to the values respective for each line at 5~K. N\'{e}el temperature is marked with the line.}
 \label{fig:Suppl.Ramanspectra_vs_T}
\end{figure*}

We register a series of Raman spectra as a function of temperature from 5~K up to 250~K with the detection of two mutually perpendicular linear polarizations, $H$ and $V$. An evolution of four selected spectra registered in the $H$ polarization is shown in Suppl.~Fig.~\ref{fig:Suppl.Ramanspectra_vs_T}a-c. To extract an intensity, energy position, and linewidth a Lorentzian curve is fitted to each of the lines. Suppl.~Fig.~\ref{fig:Suppl.Ramanspectra_vs_T}d shows intensities of the lines normalized to the intensity value at 5~K. The observed intensity dependencies are comparable for the signal registered in both linear polarizations (not shown). A general trend for all the lines is a decrease in intensity with the increasing temperature. After an initial quick drop, the decrease in intensity lowers its rate above the T$_{N} = 38$~K, marking the position of the magnetic phase transition to a paramagnetic phase. The oscillations of intensity superimposed on a general trend are discussed in the further part of the text. 

A relative shift shown in Suppl.~Fig.~\ref{fig:Suppl.Ramanspectra_vs_T}e represents a spectral position of a given line as a function of temperature after subtraction of its position at 5~K. While all lines shift towards lower energies with the increase of temperature in the full considered temperature range, anomalous behaviour is observed for the X$_3$ line below the T$_{N}$. This is consistent with the theoretical considerations predicting spectral a blueshift of the X$_3$ when a probability of other arrangements than the A-AFM increases.
The X$_4$ line (at around 409.8~cm$^{-1}$) shows a sudden change in energy upon approaching the N\'{e}el temperature. Suppl.~Fig.~\ref{fig:Suppl.Ramanspectra_vs_T}f shows the linewidth of the respective lines. Each of the lines broadens with the increase of the temperature. %In the case of the X$_2$ line a broadening is sudden while approaching the N\'{e}el temperature.

\begin{figure*}[h!]
  \includegraphics[width=0.8\textwidth]{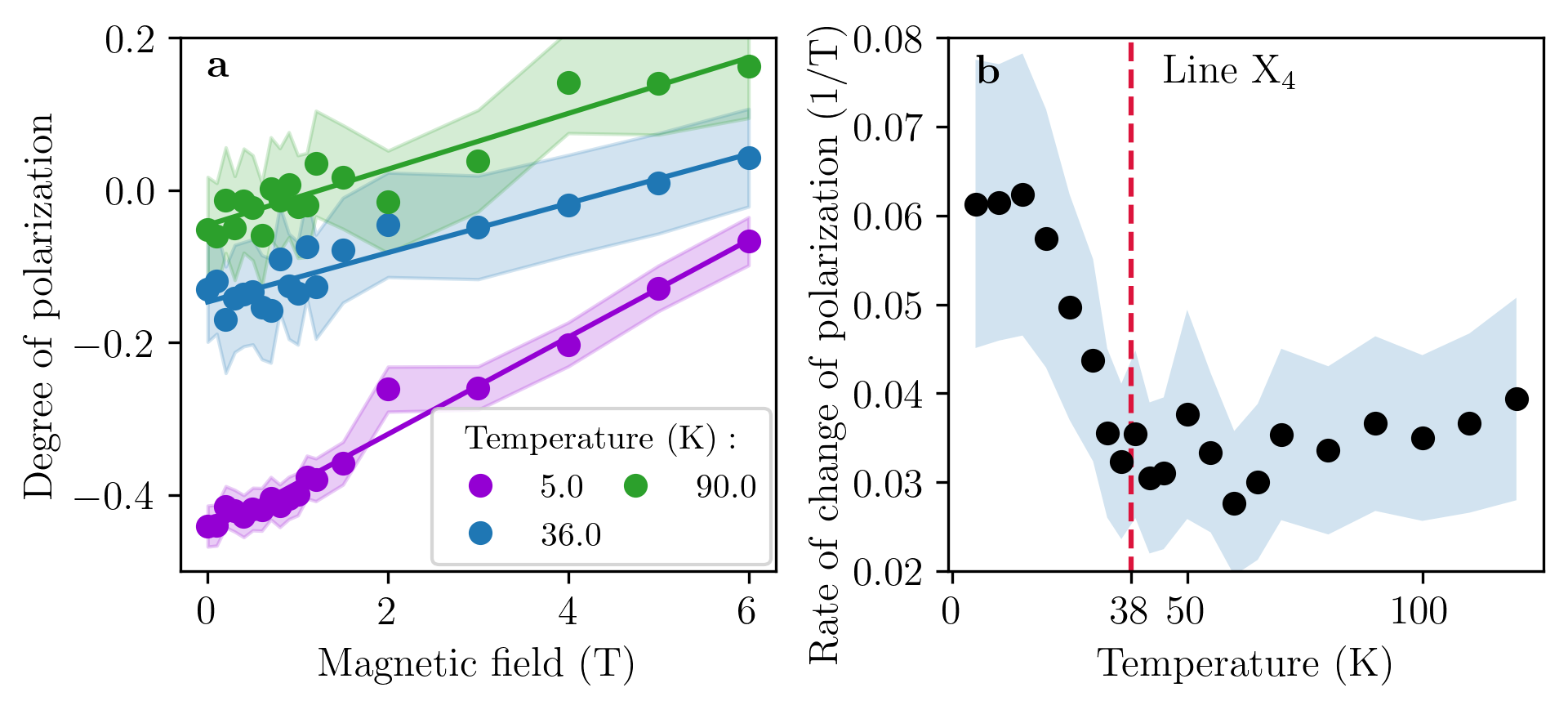}
	\caption{(a) Degree of circular polarization of the X$_4$ line in the Raman spectrum as a function of temperature and magnetic field for temperatures of 5~K, 36~K, and 90~K. (b) Slopes of the magnetic field induced change of the degree of circular polarization of the X$_4$ line as a function of temperature in the range between 5 K and 150 K.}
	\label{supplfig:Ramanpoldeg}
\end{figure*}

The results shown in Suppl. Fig.~\ref{supplfig:Ramanpoldeg} are obtained from the same set of data, which served for the preparation of Fig.~\ref{fig:Raman_vs_suscept} discussed in the main text. Suppl.~Fig.~\ref{supplfig:Ramanpoldeg}a shows a degree of circular polarization of the X$_4$ line as a function of the magnetic field for three selected temperatures. As we can see, the degree of polarization varies with the increase of the magnetic field in an approximately linear way. The slopes of the linear dependencies fitted to the data points (solid lines in Suppl.~Fig.~\ref{supplfig:Ramanpoldeg}a) are plotted as a function of temperature in Suppl.~Fig.~\ref{supplfig:Ramanpoldeg}b. The position of the N\'{e}el temperature is clearly marked on the dependence by a change of the value of the slopes, which decrease in the range approximately from 5~K to 38~K and increase slightly above the T$_N$. Presented results further confirm the usefulness of the Raman magnetospectroscopy in the studies of magnetic phase transitions in antiferromagnetic semiconductors.

\begin{figure*}[h!]
  \includegraphics[width=0.9\textwidth]{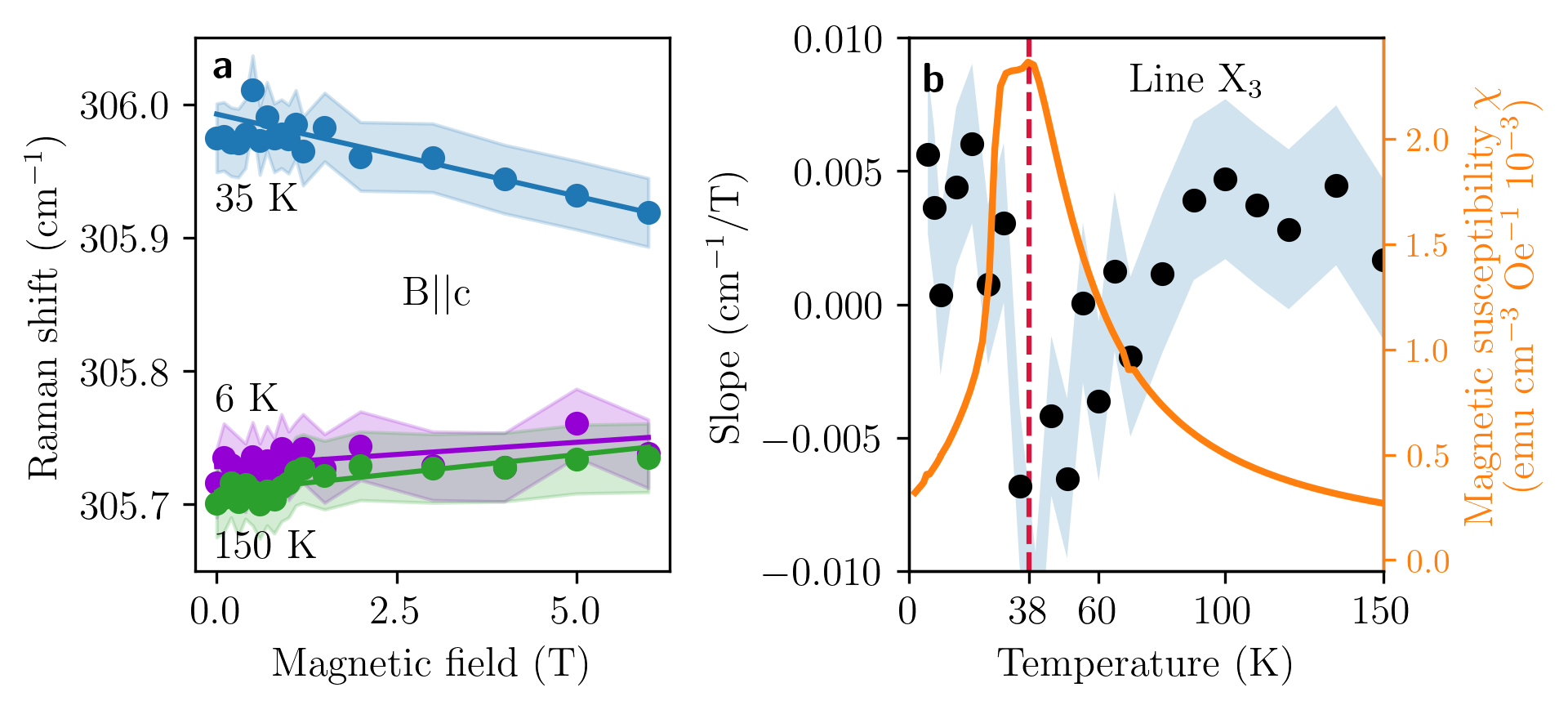}
	\caption{(a) The spectral position of the X$_3$ line in the Raman spectrum as a function of magnetic field for temperatures of 6~K, 35~K, and 150~K. A shift with the field is the most pronounced for the temperatures in the N\'{e}el temperature vicinity. (b) Slopes of the magnetic field induced shifts of X$_3$ as a function of temperature in a range between 5~K and 150~K. We note that the shift occurs in an opposite direction to the line X$_4$ discussed in the main text. The dependence exhibits the same shape as the magnetic susceptibility curve determined for the parallel configuration in SQUID, but is flipped vertically.}
	\label{fig:Raman_vs_suscepX3}
\end{figure*}

The results shown in Suppl.~Fig.~\ref{fig:Raman_vs_suscepX3} are obtained from the same measurement as those in Fig.~\ref{fig:Raman_vs_suscept} in the main text. The Suppl.~Fig.~\ref{fig:Raman_vs_suscepX3}a presents a plot of the Raman shift of line X$_3$ (305.7~cm$^{-1}$ at 5 K) as a function of applied magnetic field for selected temperatures. Also in the case of this line, the strongest spectral shift induced by magnetic field is observed in the vicinity of N\'{e}el temperature. We note, however, this shift occurs in an opposite direction with respect to that observed for the X$_4$ line discussed in Fig.~\ref{fig:Raman_vs_suscept} in the main text. Consistently, the slopes plotted in Suppl.~Fig.~\ref{fig:Raman_vs_suscepX3}b follow the negative of the magnetic susceptibility $\chi$.

\begin{figure}[h!]
\centering
\includegraphics[width=0.8\textwidth]{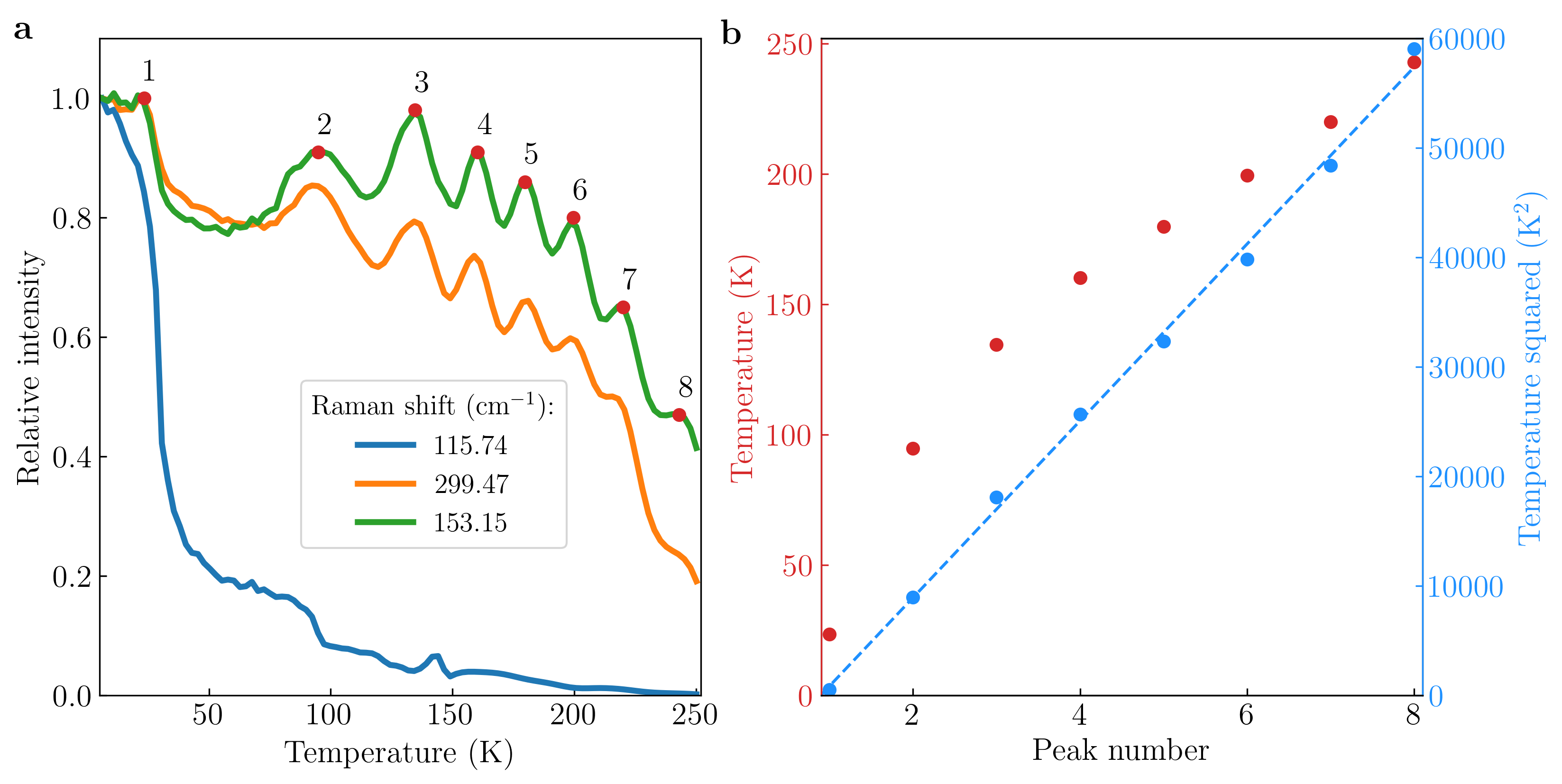} 
\caption{(a) The intensity of the three selected lines in the Raman spectrum of CrPS$_4$ as a function of temperature (positions of the lines at 5~K provided in the legend). Oscillations of the intensity are evidenced in the case of two lines. The maxima of intensity occur at the same temperature for these lines. (b) The temperature corresponding to each intensity maximum from panel (a) (red points) and the square of these values (blue points) plotted as a function of the order number of the maximum. The plot proves that values of the temperatures for which the maxima occur follow the square root of their order number: $T \sim \sqrt{n}$.}
 \label{Supplfig:oscillations}
\end{figure}

In Suppl.~Fig.~\ref{Supplfig:oscillations}a we plot the intensity of three selected lines of the Raman spectrum of CrPS$_4$ as a function of temperature. The line between the experimental points is approximated with a spline. As we can see, oscillations of the intensity superimposed in the dependence are evidenced in the case of two lines. Temperature positions of maxima of the dependence (indicated by red dots) are plotted as a function of the order number of the maximum in Suppl.~Fig.~\ref{Supplfig:oscillations}b. We can see that the temperature position of the maximum is related to the maximum number by a square root function, $T \sim \sqrt{n}$ (red points). To further solidify this claim, we plot also a square of the position in temperature as a function of the maximum number (blue points in Suppl.~Fig.~\ref{Supplfig:oscillations}b). We see that those follow a straight line. The observed result suggests the presence of a kind of temperature-dependent, resonant effect in excitation of the selected transitions in the Raman spectrum, but its interpretation requires further analysis.

\section{Theoretical modelling}
\label{SuplSec:modelling}

\begin{figure*}
\centering
\includegraphics[width=0.95\textwidth]{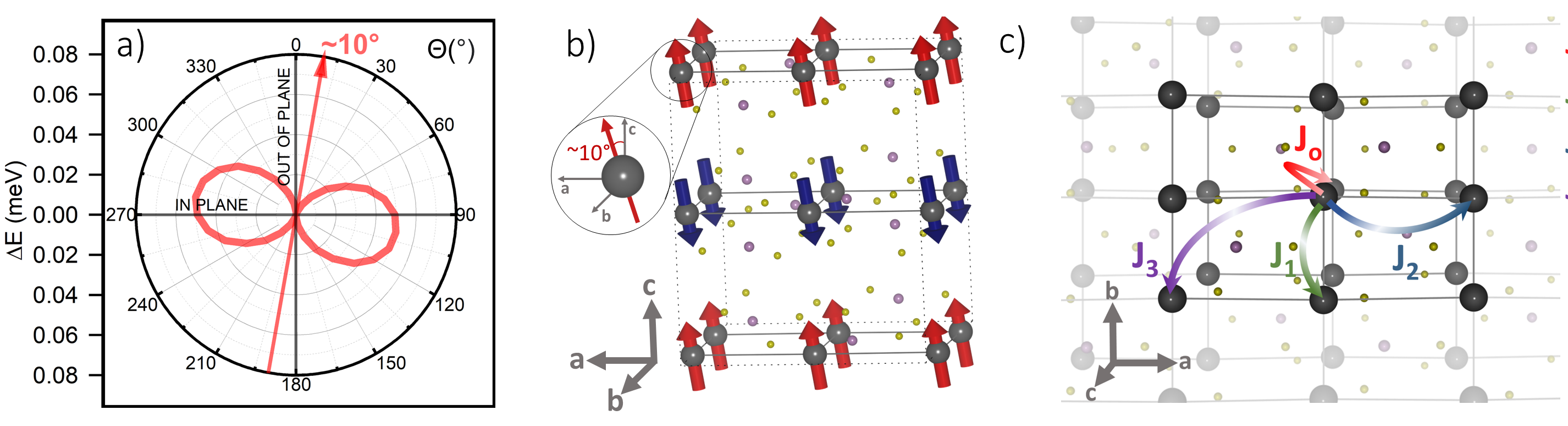}
\caption{(a) Polar plot of energy difference $\Delta$E between the particular direction of the Cr spins and the lowest energy, that is A-AFM configuration. (b) The magnetic ground state configuration is A-AFM phase with the magnetic easy axis tilted 10$^o$ from the $c$-axis. (c) Schematic picture of the exchange coupling strengths between 1NN (J$_1$), 2NN (J$_2$) and 3NN (J$_3$), as well as out-of-plane exchange coupling (J$_0$).}
\label{Supplfig:Supp_exch}
\end{figure*}

\textit{Magnetocrystalline energy}. We start our calculations by examining the magnetocrystalline energy (MAE), including the spin-orbit coupling (SOC) within the DFT+U approach. The MAE is defined here, as a difference between the energies of spins parallel to hard (highest energy) and easy axes (lowest energy) of magnetization. We rotate the spins from out-of-plane to in-plane configurations with an angle step equal to $10^{o}$. The energy results are presented in the polar plot shown in Suppl.~ Fig.~\ref{Supplfig:Supp_exch}a. The MAE is equal to 0.05~meV per Cr$^{3+}$ ion (0.4 meV per a bulk supercell containing 8~Cr atoms), which is a similar value like obtained for other 2D magnetic crystals such as MnPSe$_3$ and NiPSe$_3$ as reported recently.\cite{PhysRevB.109.054426} The magnetic easy axis is tilted $10^{o}$ from the crystallographic \textit{c}-axis, and the magnetic ground state is A-AFM presented in Suppl.~Fig.~\ref{Supplfig:Supp_exch}b. Due to the tiny difference between the energies of the spins lying along the hard and easy axes, we  assume only spin-polarized case of the Cr$^{3+}$ ions in further considerations.

\textit{Magnetic interactions}. The exchange coupling constants have been derived from the classical spin Heisenberg model in a similar manner, as described in Ref.\cite{Autieri2022}. To achieve this, we fix the lattice parameters and position of the atoms to the magnetic ground state (A-AFM), and we calculate the total energy for five different magnetic arrangements within the spin-polarized case of DFT+U assuming the same lattice parameters and position of the atoms. Next, these energies are mapped onto the effective spin Heisenberg model, which enables us to obtain the nearest neighbor (NN) exchange coupling constants collected in Suppl.~Table~\ref{Suppltable:exchconst}. The obtained values are in good agreement with the literature.\cite{Calder:PhysRevB2020}

\begin{table}
 \def\arraystretch{1.5}
\begin{center}
\begin{tabular}{  c c  }
 J$_{ij}$ (d$_{ij}$)& \\
     \hline
$J_1$ ($d_1$=3.59\r{A}) &2.131 meV \\
$J_2$ ($d_2$=5.44\r{A})   &0.496 meV \\ 
$J_3$ ($d_3$=6.52\r{A})   &0.248 meV \\ 
$J_o$ ($d_0$=6.20\r{A})   &-0.141 meV \\ 
\hline
\end{tabular}
\caption{\label{Suppltable:exchconst}Exchange coupling strengths between 1NN (J$_1$), 2NN (J$_2$) and 3NN (J$_3$) nearest neighbors, as well as  out-of-plane nearest neighbors (J$_0$). Negative and positive $J_{i}$ values denote FM and AFM couplings, respectively. } 
\end{center}
\end{table}

\begin{figure*}[h!]
\centering
\includegraphics[width=0.7\textwidth]{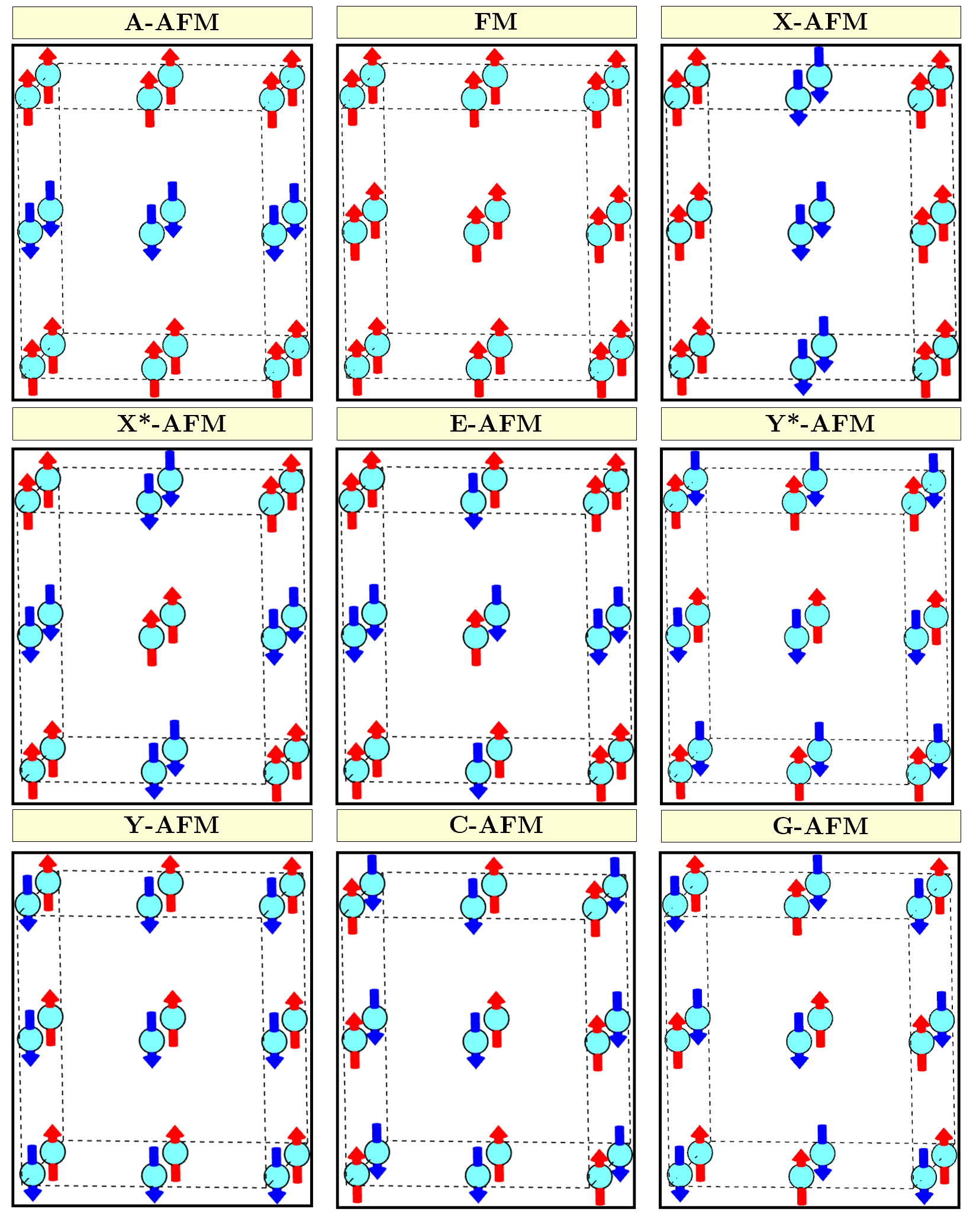}
\caption{Various spin arrangements of Cr$^{3+}$ sites employed in this study. The dotted lines indicate a supercell used (Cr$_8$P$_8$S$_{32}$).}
\label{Supplfig:spinarr}
\end{figure*}

\textit{Magnetic phases}. We consider all non-equivalent spin arrangements for the employed supercell within DFT+U (see Suppl.~Fig.~\ref{Supplfig:spinarr}). For each of the spin configurations, the lattice parameters and position of atoms have been fully optimized and collected in Suppl.~Table \ref{Suppltab:magorders_energy}. In the view of the calculations, the A-AFM phase is the lowest-energy state followed closely by the FM phase. The energy difference between these two phases is around 1~meV per each magnetic ion. The next lowest phase, X-AFM lies 12~meV above the energy minimum set by the A-AFM phase, whereas the highest-energy phase is G-AFM, with the energy around 26~meV per magnetic ion. %Thus, at T~=~38~K (k$_B(\mathrm{T}=38 K)=3.27$~meV), only A-AFM and FM phases may be expected to co-exist through thermally-activated spin flip transitions.

Notably, going from the A-AFM to the Y-AFM phase (17~meV above A-AFM) involves a transition between out-of-plane to in-plane antiferromagnetic arrangements. Compared to a fully ferromagnetic system, the A-AFM phase represents the activation of out-of-plane antiferromagnetism with in-plane ferromagnetism preserved, while the opposite is true for Y-AFM~--~it remains ferromagnetic with respect to the nearest out-of-plane neighbors, while the nearest-neighbor interaction is antiferromagnetic in the in-plane direction (see Suppl.~Fig.~\ref{Supplfig:spinarr}). This observation makes it interesting to compare the properties of these three systems: the ferromagnetic phase as a reference, and the two AFM phase as prototype out-of-plane and in-plane antiferromagnets. Tracking the temperature evolution of the simulated Raman modes allows us to partially decouple the influence of structural and magnetic degrees of freedom in CrPS$_4$. This can be achieved by analyzing which vibration modes are sensitive to the change in magnetic order, and which are influenced mostly by structural changes in the form of simulated temperature increase.

\begin{figure*}[h!]
\centering
\includegraphics[width=0.5\textwidth]{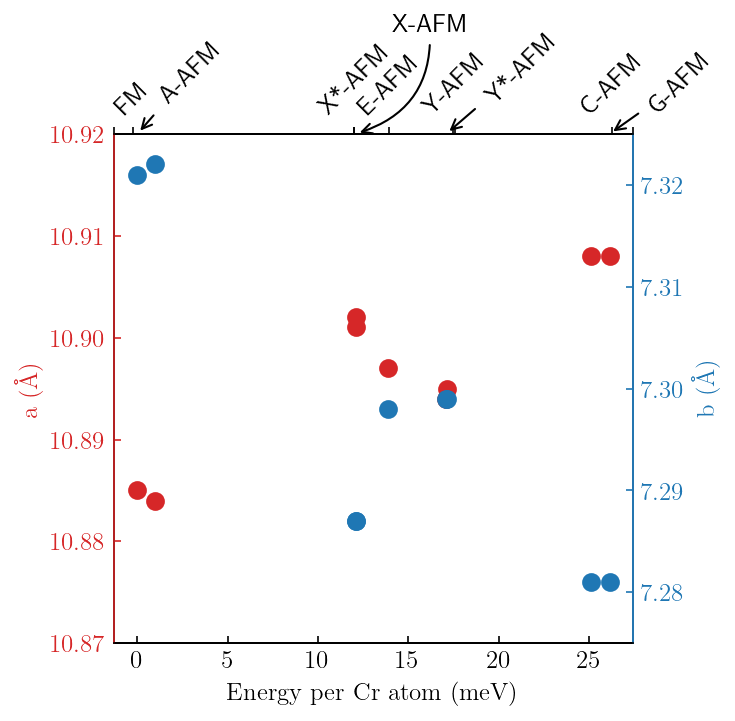}
\caption{Calculated optimal lattice parameters $a$ and $b$ of CrPS$_4$ as a function of relative energy difference of various magnetic arrangements of Cr$^{3+}$ spins (indicated in the top horizontal axis) with respect to the magnetic A-AFM groundstate (bottom horizontal axis). The plot shows the data collected in Suppl.~Table~\ref{Suppltab:magorders_energy}. The presented results indicate that the lattice parameter $a$ and $b$ respectively enlarges and diminishes with the increasing energy per Cr atom (hence with the increasing temperature).}
\label{supplfig:ablattices_theor}
\end{figure*}

\textit{Impact of structural changes determined in PXRD measurements on Raman modes.} Finally, we examine how the changes in the lattice parameters obtained by  PXRD  measurements affect the optically active Raman modes. In this regard, we used DFT+U approach and we took into account directly the changes of the lattice parameters for a given temperature in accordance  to the strategy described below. First, geometry relaxation was performed for the 8-formula-unit supercell with different magnetic arrangements presented in Suppl.~Fig.~\ref{Supplfig:spinarr} and the results of the optimized lattice parameters for each employed spin configuration are collected in the Suppl.~Table~\ref{Suppltab:magorders_energy} and Suppl.~Fig.~\ref{supplfig:ablattices_theor}. 

The data collected in Suppl.~Table~\ref{Suppltab:magorders_energy} for $a$ and $b$ lattice parameters of CrPS$_4$ are plotted in Suppl.~Fig.~\ref{supplfig:ablattices_theor}. The plot shows values of the $a$ and $b$ lattice parameters calculated as a function of a relative energy difference $\Delta E$ per Cr atom with respect to the magnetic A-AFM groundstate (bottom horizontal axis). Corresponding magnetic arrangements for respective $\Delta E$ values are indicated in the top horizontal axis of the plot. Taking into account that the probability of deviation from A-AFM state and thus of transition to a higher energy magnetic arrangement increases with the increasing temperature, the calculation predicts in an accurate way the evolution of the $a$ and $b$ lattice parameters with the temperature observed in the XRD experiment reported in the main text.

\begin{table*}[h!]
\centering
\resizebox{\textwidth}{!}{%
\begin{tabular}{cc|c|c|c|c|c|c|c}
\hline
\multicolumn{2}{c|}{Magnetic phase}                 & $\Delta E$ & a (Å)  & b (Å) & c (Å) & $\alpha$ (deg) & $\beta$ (deg) & $\gamma$ (deg) \\ \hline
\multicolumn{1}{c|}{\multirow{2}{*}{(I)}}   & A-AFM ($d_1$=3.589 \r{A})  & 0          & 10.885 & 7.321 & 6.202 & 90.00          & 92.38         & 90.00          \\
\multicolumn{1}{c|}{}                       & FM ($d_1$=3.590 \r{A})    & 0.98       & 10.884 & 7.322 & 6.202 & 90.00          & 92.39         & 90.00          \\ \hline
\multicolumn{1}{c|}{\multirow{5}{*}{(II)}}  & X-AFM ($d_1$=3.588 \r{A})  & 12.08      & 10.902 & 7.287 & 6.217 & 90.00          & 92.19         & 90.00          \\ 
\multicolumn{1}{c|}{}                       & X*-AFM ($d_1$=3.587 \r{A}) & 12.09      & 10.901 & 7.287 & 6.217 & 90.00          & 92.19         & 90.00          \\
\multicolumn{1}{c|}{}                       & E-AFM ($d_1$=3.572 \r{A})  & 13.86      & 10.897 & 7.298 & 6.208 & 90.00          & 92.26         & 90.00          \\
\multicolumn{1}{c|}{}                       & Y*-AFM ($d_1$=3.573 \r{A}) & 17.09      & 10.894 & 7.299 & 6.201 & 90.00          & 92.30         & 90.00          \\
\multicolumn{1}{c|}{}                       & Y-AFM  ($d_1$=3.573 \r{A}) & 17.15      & 10.895 & 7.299 & 6.202 & 90.00          & 92.30         & 90.00          \\ \hline
\multicolumn{1}{c|}{\multirow{2}{*}{(III)}} & C-AFM  ($d_1$=3.573 \r{A}) & 25.09      & 10.908 & 7.281 & 6.207 & 90.00          & 92.16         & 90.00          \\
\multicolumn{1}{c|}{}                       & G-AFM  ($d_1$=3.574 \r{A}) & 26.14      & 10.908 & 7.281 & 6.21  & 90.00          & 92.17         & 90.00          \\ \hline
\end{tabular}
}
\caption{Optimal lattice parameters for various magnetic arrangements (see Suppl.~Fig.~\ref{Supplfig:spinarr}) determined by DFT+U calculations. Energy difference $\Delta E$ calculated with respect to the magnetic ground state is given in meV per Cr$^{3+}$ ion. The (I), (II) and (III) indicate the type of the in-plane coupling between the Cr$^{3+}$ spins: FM, mixed AFM and FM, and AFM couplings (only the first and second nearest neighbour considered), respectively.}
\label{Suppltab:magorders_energy}
\end{table*}

\begin{figure*}[h!]
\centering
\includegraphics[width=\textwidth]{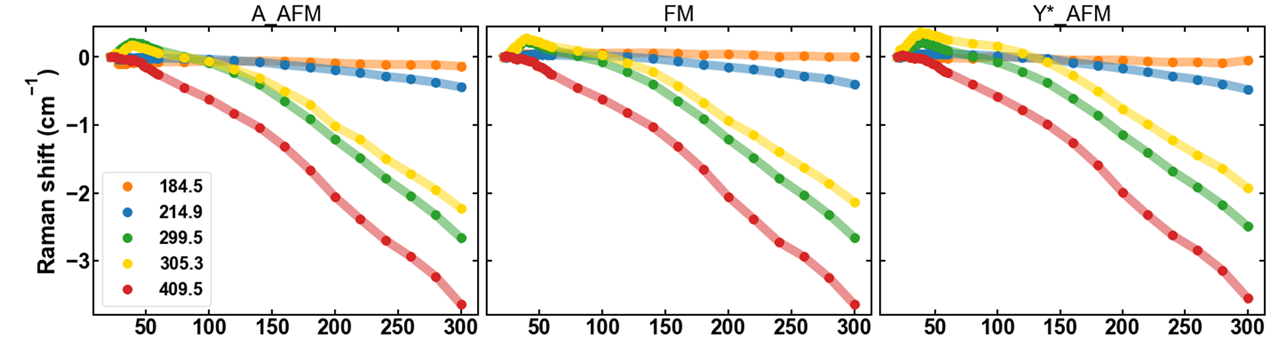}
\caption{The calculated impact of the temperature-induced lattice parameters variation on the frequency of Raman modes (with a spectral position in cm$^{-1}$ at 5~K provided in the legend) of CrPS$_4$. Raman shift denotes the difference between the phonon frequencies calculated for a particular temperature and phonon frequencies obtained for the optimized system (e.g., for FM phase: (X$^{FM}_T$ - X$^{FM}_{20 K}$)). Note, that for each magnetic phase, a different optimized system might be obtained.}
\label{Supplfig:teorshiftvsT}
\end{figure*}

Then, temperature effects are taken into account by scaling the $a, b, c$, and $\beta$ parameters according to the data obtained from temperature-dependent PXRD measurements. Each arrangement is initially allowed to relax to its ground state, before applying a dimensionless temperature factor in the form of a$^i_T$/a$^i_{20K}$, where a$^i_T$ is the value lattice parameter a$^i$ (where i=1,2,3, a$^1$=a, a$^2$=b, a$^3$=c) at temperature T, and a$^i_{20K}$ is the value of parameter a$^i$ at 20~K, the lowest temperature considered in the PXRD measurements. In other words, the temperature is taken indirectly into consideration, by the factor defined by the lattice parameters. Note that temperature evolution of the optically active Raman modes for the three prototype magnetic phases presented in Suppl.~Fig.~\ref{Supplfig:teorshiftvsT} exhibits a qualitatively similar trend for the full temperature range.  Thus, the evolution of Raman shifts (Raman shift denotes here a difference between the frequencies at the given temperature and ground state (lowest temperature for the same magnetic arrangement) is rather independent of magnetic ordering, and reflects structural changes in lattice parameters. A direct comparison between the corresponding frequencies of the Raman modes in CrPS$_4$ for various magnetic phases is collected in Suppl.~Table~\ref{Suppltab:phonons}.

\newcolumntype{Y}{>{\centering\arraybackslash}X}
\newcolumntype{L}{>{\flushleft\arraybackslash}X}
\newcolumntype{R}{>{\flushright\arraybackslash}X}

\begin{table*}[b]
\caption{Experimental and calculated Raman mode frequencies of bulk CrPS$_4$. Mode frequencies are calculated for A-AFM ground state, as well as for FM, Y*-AFM and G-AFM magnetic arrangements. A shift with respect to the A-AFM ground state is provided for FM, Y*-AFM and G-AFM arrangements. Blue- and red-shifted modes are indicated by the font color. Bold formatting of the text highlights the modes with a significant (above 0.1~cm$^{-1}$) shift from A-AFM. The first column provides information of the irreducible representation of the mode. The X$_1$, X$_2$, X$_3$ and X$_4$ transitions considered in the main text are indicated.}
\label{Suppltab:phonons}
\begin{tabularx}{\textwidth}{R|R|R|RR|RR|RR}
\hline
mode & experiment & A-AFM & FM & shift from A-AFM        & Y*-AFM & shift from A-AFM        & G-AFM & shift from A-AFM         \\
\hline
$B$                      & 82.90                    & 82.82                     & 82.63                  & {\textcolor[HTML]{FF0000} {\ul \textbf{-0.19}}} & 82.94                      & {\textcolor[HTML]{00B0F0} {\ul \textbf{0.12}}}  & 82.01                     & {\textcolor[HTML]{FF0000} {\ul \textbf{-0.81}}}  \\
$B$                         & 113.20                   & 110.95                    & 110.79                 & {\textcolor[HTML]{FF0000} {\ul \textbf{-0.16}}} & 111.64                     & {\textcolor[HTML]{00B0F0} {\ul \textbf{0.69}}}  & 111.65                    & {\textcolor[HTML]{00B0F0} {\ul \textbf{0.70}}}   \\
$A$                         & 116.10                   & 117.07                    & 117.01                 & {\textcolor[HTML]{FF0000} {-0.06}}                & 116.62                     & {\textcolor[HTML]{FF0000} {\ul \textbf{-0.46}}} & 115.94                    & {\textcolor[HTML]{FF0000} {\ul \textbf{-1.14}}}  \\
$B$                         & 153.60                   & 145.24                    & 145.42                 & {\textcolor[HTML]{00B0F0} {\ul \textbf{0.18}}}  & 147.93                     & {\textcolor[HTML]{00B0F0} {\ul \textbf{2.69}}}  & 147.70                    & {\textcolor[HTML]{00B0F0} {\ul \textbf{2.45}}}   \\
$A$                         & 169.80                   & 164.99                    & 165.24                 & {\textcolor[HTML]{00B0F0} {\ul \textbf{0.24}}}  & 161.52                     & {\textcolor[HTML]{FF0000} {\ul \textbf{-3.47}}} & 152.06                    & {\textcolor[HTML]{FF0000} {\ul \textbf{-12.94}}} \\
(X$_{1}$) $A$                        & 184.50                   & 180.38                    & 180.37                 & {\textcolor[HTML]{FF0000} {-0.01}}                & 171.47                     & {\textcolor[HTML]{FF0000} {\ul \textbf{-8.91}}} & 167.09                    & {\textcolor[HTML]{FF0000} {\ul \textbf{-13.29}}} \\ 
$A$                         & 215.40                    & 208.19                    & 208.32                 & {\textcolor[HTML]{00B0F0} {\ul \textbf{0.13}}}  & 208.38                     & {\textcolor[HTML]{00B0F0} {\ul \textbf{0.18}}}  & 207.75                    & {\textcolor[HTML]{FF0000} {\ul \textbf{-0.44}}}  \\
$B$                         & 231.60                   & 223.13                    & 223.03                 & {\textcolor[HTML]{FF0000} {-0.10}}                & 223.88                     & {\textcolor[HTML]{00B0F0} {\ul \textbf{0.75}}}  & 222.44                    & {\textcolor[HTML]{FF0000} {\ul \textbf{-0.69}}}  \\
$B$                         & 257.50                   & 243.56                    & 243.48                 & {\textcolor[HTML]{FF0000} {-0.08}}                & 245.44                     & {\textcolor[HTML]{00B0F0} {\ul \textbf{1.88}}}  & 245.50                    & {\textcolor[HTML]{00B0F0} {\ul \textbf{1.94}}}   \\
$A$                         & 269.40                   & 257.71                    & 257.57                 & {\textcolor[HTML]{FF0000} {\ul \textbf{-0.14}}} & 255.51                     & {\textcolor[HTML]{FF0000} {\ul \textbf{-2.20}}} & 255.01                    & {\textcolor[HTML]{FF0000} {\ul \textbf{-2.70}}}  \\ 
(X$_{2}$) $B$                & 299.50                    & 286.47                    & 286.35                 & {\textcolor[HTML]{FF0000} {\ul \textbf{-0.12}}} & 286.52                     & {\textcolor[HTML]{00B0F0} {0.05}}                 & 285.23                    & {\textcolor[HTML]{FF0000} {\ul \textbf{-1.25}}}  \\ 
(X$_{3}$) $A$              & 305.70                     & 289.74                    & 289.73                 & {\textcolor[HTML]{FF0000} {-0.01}}                & 290.95                     & {\textcolor[HTML]{00B0F0} {\ul \textbf{1.22}}}  & 290.19                    & {\textcolor[HTML]{00B0F0} {\ul \textbf{0.46}}}   \\
$A$                         & 324.50                   & 310.55                    & 310.80                 & {\textcolor[HTML]{00B0F0} {\ul \textbf{0.25}}}  & 308.63                     & {\textcolor[HTML]{FF0000} {\ul \textbf{-1.91}}} & 309.30                    & {\textcolor[HTML]{FF0000} {\ul \textbf{-1.25}}}  \\
$B$                         & 350.30                   & 335.20                    & 335.27                 & {\textcolor[HTML]{00B0F0} {0.07}}                 & 336.17                     & {\textcolor[HTML]{00B0F0} {\ul \textbf{0.97}}}  & 337.42                    & {\textcolor[HTML]{00B0F0} {\ul \textbf{2.22}}}   \\ 
(X$_{4}$) $A$                & 409.80                    & 393.09                    & 393.17                 & {\textcolor[HTML]{00B0F0} {0.08}}                 & 390.90                     & {\textcolor[HTML]{FF0000} {\ul \textbf{-2.19}}} & 391.96                    & {\textcolor[HTML]{FF0000} {\ul \textbf{-1.13}}}  \\
$A$                         & 523.20                   & 497.33                    & 497.33                 & {\textcolor[HTML]{FF0000} {-0.01}}                & 495.40                     & {\textcolor[HTML]{FF0000} {\ul \textbf{-1.94}}} & 496.36                    & {\textcolor[HTML]{FF0000} {\ul \textbf{-0.97}}}  \\
$A$                         & 543.10                   & 526.09                    & 526.23                 & {\textcolor[HTML]{00B0F0} {\ul \textbf{0.14}}}  & 527.49                     & {\textcolor[HTML]{00B0F0} {\ul \textbf{1.40}}}  & 527.31                    & {\textcolor[HTML]{00B0F0} {\ul \textbf{1.22}}}   \\
$B$           &       -             & 575.11                    & 575.37                 & {\textcolor[HTML]{00B0F0} {\ul \textbf{0.26}}}  & 573.88                     & {\textcolor[HTML]{FF0000} {\ul \textbf{-1.23}}} & 573.18                    & {\textcolor[HTML]{FF0000} {\ul \textbf{-1.93}}} \\
\hline
\end{tabularx}
\end{table*}

\end{document}